\documentclass[twocolumn]{aastex701}

\received{December 12, 2025}
\revised{February 24, 2026}
\accepted{\today}

\submitjournal{ApJ}

\begin{document}

\title{\textit{Spitzer} + HST parallaxes of 13 late T and Y dwarfs.}

\correspondingauthor{Federico Marocco}
\email{federico@ipac.caltech.edu}

\author[0000-0001-7519-1700]{Federico Marocco}
\affiliation{IPAC, Mail Code 100-22, Caltech, 1200 E. California Blvd., Pasadena, CA 91125, USA}
\email{federico@ipac.caltech.edu}

\author[0000-0003-4269-260X]{J. Davy Kirkpatrick}
\affiliation{IPAC, Mail Code 100-22, Caltech, 1200 E. California Blvd., Pasadena, CA 91125, USA}
\email{davy@ipac.caltech.edu}

\author[0000-0002-4424-4766]{Richard L. Smart}
\affiliation{Istituto Nazionale di Astrofisica, Osservatorio Astrofisico di Torino, Strada Osservatorio 20, I-10025 Pino Torinese, Italy}
\email{}

\author[0000-0002-6294-5937]{Adam C. Schneider}
\affiliation{United States Naval Observatory, Flagstaff Station, 10391 West Naval Observatory Rd., Flagstaff, AZ 86005, USA}
\email{}

\author[0000-0001-7896-5791]{Dan Caselden}
\affiliation{Department of Astrophysics, American Museum of Natural History, Central Park West at 79th Street, New York, NY 10024, USA}
\email{}

\author[0000-0003-4142-1082]{Edgardo Costa}
\affiliation{Astronomy Department, Universidad de Chile, Casilla 36-D, Santiago, Chile}
\email{}

\author[0000-0001-7780-3352]{Michael C. Cushing}
\affiliation{Ritter Astrophysical Research Center, Department of Physics \& Astronomy, University of Toledo, 2801 W. Bancroft St., Toledo, OH 43606, USA}
\email{}

\author[0009-0000-5181-7924]{Maximiliano Dirk}
\affiliation{Centre for Astrophysics Research, University of Hertfordshire, College Lane, Hatfield AL10 9AB, UK}
\affiliation{Istituto Nazionale di Astrofisica, Osservatorio Astrofisico di Torino, Strada Osservatorio 20, I-10025 Pino Torinese, Italy}
\email{}

\author{Peter R. M. Eisenhardt}
\affiliation{Jet Propulsion Laboratory, California Institute of Technology, 4800 Oak Grove Dr., Pasadena, CA 91109, USA}
\email{}

\author[0000-0001-6251-0573]{Jacqueline K. Faherty}
\affiliation{Department of Astrophysics, American Museum of Natural History, Central Park West at 79th Street, New York, NY 10024, USA}
\email{}

\author[0000-0001-5072-4574]{Christopher R. Gelino}
\affiliation{IPAC, Mail Code 100-22, Caltech, 1200 E. California Blvd., Pasadena, CA 91125, USA}
\email{}

\author[0000-0002-2387-5489]{Marc J. Kuchner}
\affiliation{Exoplanets and Stellar Astrophysics Laboratory, NASA Goddard Space Flight Center, 8800 Greenbelt Road, Greenbelt, MD 20771, USA}
\email{}

\author[0000-0002-1125-7384]{Aaron M. Meisner}
\affiliation{NSF’s National Optical-Infrared Astronomy Research Laboratory, 950 N. Cherry Avenue, Tucson, AZ 85719, USA}
\email{}

\author[0000-0003-1454-0596]{Rene A. Mendez}
\affiliation{Astronomy Department, Universidad de Chile, Casilla 36-D, Santiago, Chile}
\email{}

\author[0000-0002-7181-2554]{Robert A. Stiller}
\affiliation{Ritter Astrophysical Research Center, Department of Physics \& Astronomy, University of Toledo, 2801 W. Bancroft St., Toledo, OH 43606, USA}
\email{}

\author[0000-0001-5058-1593]{Edward L. Wright}
\affiliation{Department of Physics and Astronomy, University of California Los Angeles, 430 Portola Plaza, Box 951547, Los Angeles, CA 90095-1547, USA}
\email{}

\begin{abstract}

We present astrometric measurements for 13 cold brown dwarfs in the solar neighborhood ($d < 20$pc). By combining archival \textit{Spitzer} data with our own Hubble Space Telescope (HST) observations, we achieve parallax uncertainties typically around 10\%. Using \textit{Spitzer} and HST photometry we compare our targets with other known late T and Y dwarfs in the Solar neighborhood, confirming that there is large intrinsic scatter in the near- and mid-infrared absolute magnitudes and colors of this population, further highlighting the diversity observed spectroscopically by several James Webb Space Telescope (JWST) programs. This scatter makes photometric distance estimates highly unreliable and, therefore, makes astrometric parallax measurements fundamental for a meaningful characterization of even the nearest cold brown dwarfs.

\end{abstract}


\section{Introduction} \label{sec:intro}

The selection of volume-limited, complete samples is a necessary starting point for the characterization of celestial objects, and for the understanding of the physical processes underlying fundamental astrophysical phenomena. One such example is the determination of the initial mass function \citep[hereafter IMF;][]{1955ApJ...121..161S}, which is one of the most important observation-based anchors for star formation theory. While, in principle, the IMF can be determined from a magnitude-limited sample, several observational and statistical biases (e.g. the Lutz-Kelker effect; \citealt{1973PASP...85..573L}) will plague the results. Similarly, the sample must be complete across the mass range over which the IMF is being evaluated, and the completeness must be carefully assessed \citep[see e.g.][]{1968ApJ...151..393S}.

\citet{2024ApJS..271...55K} assembled the most complete census of the stellar and substellar occupants of the spherical volume centered on the Solar system and extending out to 20 pc. The majority of objects in their census are securely placed within such volume by precise astrometric distance measurements, derived from \textit{Gaia} DR3 \citep{2023A&A...674A...1G} or, for many substellar objects, obtained through dedicated observing campaigns \citep[e.g.][]{2021ApJS..253....7K,2021AJ....161...42B,2018MNRAS.481.3548S,2017AJ....153...14W}. The membership of several sources, however, is only tentatively established via spectro-photometric distance estimates. This is particularly true for the coldest, lowest-mass members of the 20pc census -- brown dwarfs. These substellar objects are too faint at optical wavelenghts to be observed by \textit{Gaia}, and are challenging even for large ground-based infrared telescopes. The use of photometric distance estimates can bias the space density measurements (for example through the incorrect inclusion of unrecognized unresolved binaries in the sample), in turn affecting our ability to draw firm conclusions on the underlying mass function. Cold brown dwarfs sample the lowest mass end of the mass function, and are, therefore, crucial to empirically establish its cutoff \citep{2024ApJS..271...55K} . 

The \textit{Spitzer} telescope has provided parallaxes for some of the coldest nearby objects \citep{2021ApJS..253....7K,2013Sci...341.1492D}, but more substellar neighbors have been discovered close to or after the end of the \textit{Spitzer} mission. Several archival datasets, as well as past and ongoing large-area surveys, are capable of detecting these objects, but when taken individually they either provide too few observations for robust astrometric measurements, cover too short a time span to disentangle proper motion from (apparent) parallactic motion, or lack the astrometric precision to achieve meaningful distance and motion accuracy. For example, the WISE/NEOWISE \citep{2010AJ....140.1868W,2014ApJ...792...30M} dataset spans nearly 15 years, but with a PSF FWHM$\sim$6\arcsec\ and 2\farcs75 pixels does not allow for accurate parallax determinations, except for the closest, brightest objects \citep{2018ApJ...862..173T}. \citet{2021ApJS..253....7K} conducted a large astrometric follow-up campaign for brown dwarfs, supplementing the WISE/NEOWISE data with dedicated \textit{Spitzer} observations, but the end of the \textit{Spitzer} mission left several targets in their observing program with too few measurements to obtain reliable distances. 

In this paper we present further follow-up for 13 targets from the \citet{2021ApJS..253....7K} program, using the Wide Field Camera 3 \citep[WFC3;][]{2008SPIE.7010E..1EK} on HST. Section~\ref{sec:selection} describes our target selection; in Section~\ref{sec:data} we discuss our observing strategy, and in Section~\ref{sec:astrometry} we give details of our data reduction procedure; in Section~\ref{sec:results} we present the results of our astrometric measurements and put the sample in context with the rest of the nearby substellar population; finally, in Section~\ref{sec:conclusions} we summarize our findings.

\section{Target selection} \label{sec:selection}

The 13 targets of this paper were originally part of a large \textit{Spitzer} astrometric campaign to measure parallaxes for all L, T, and Y dwarfs suspected to be within 20pc \citep{2019ApJS..240...19K,2021ApJS..253....7K}. The details of the target selection are given in \citet{2019ApJS..240...19K}. With the end of the \textit{Spitzer} mission several targets were left with incomplete data sets. Those observations either covered only one of the two vertices of the parallactic ellipse (therefore rendering a measurement of the parallax impossible), or covered one vertex too sparsely for a robust parallax measurement.

All objects in the \citet{2019ApJS..240...19K,2021ApJS..253....7K} sample have publicly available WISE/NEOWISE observations, but as mentioned in Section~\ref{sec:intro} these data lack the astrometric precision necessary to complement the \textit{Spitzer} campaign. Dedicated observations are necessary to complete the parallax measurements.

Bright targets ($J < 21.0$ mag) are being observed with ground-based facilities, and will be the subject of other papers. The 13 faintest objects ($J \geq 21.0$ mag), whose observations would be impractical from the ground, were observed with HST/WFC3. The HST targets are listed in Table~\ref{tab:targets}.

\movetabledown=70mm
\begin{rotatetable*}
\begin{deluxetable*}{l c c c c c c c c c c c}
    \tablehead{\colhead{Target ID} & \colhead{R.A.} & \colhead{decl.} & \colhead{$J_{\rm MKO}$} & \colhead{$J$ Ref.} & \colhead{W1} & \colhead{W2} & \colhead{ch1} & \colhead{ch2} & \colhead{Spitzer Ref.} & \colhead{Sp. Type} & \colhead{Type Ref.} \\ 
    \colhead{} & \colhead{hh mm ss.ss} & \colhead{dd mm ss.s} & \colhead{mag} & \colhead{} & \colhead{mag} & \colhead{mag} & \colhead{mag} & \colhead{mag} & \colhead{} & \colhead{} & \colhead{}}
    
    \tablecaption{The HST/WFC3 target list. \label{tab:targets}}

    \startdata
    CWISEP J023842.60$-$133210.7 & 02 38 42.59 & $-$13 32 11.1 & $>$21.46       & 1 & $>$19.653        & 16.324$\pm$0.083 & 19.058$\pm$0.219 & 16.329$\pm$0.024 & 1 & ($\geq$Y1) & 1 \\
    CWISEP J040235.55$-$265145.4 & 04 02 35.60 & $-$26 51 45.4 & $>$20.34       & 1 & 17.995$\pm$0.112 & 15.583$\pm$0.041 & 18.173$\pm$0.150 & 15.453$\pm$0.021 & 1 & ($\geq$Y1) & 1 \\
    CWISEP J085938.95+534908.7   & 08 59 38.92 & +53 49 08.4   & $>$19.70       & 1 & $>$19.757        & 16.009$\pm$0.061 & 18.512$\pm$0.171 & 15.998$\pm$0.024 & 1 & (Y0)       & 1 \\
    CWISEP J093852.89+063440.6   & 09 38 52.92 & +06 34 40.2   & 21.03$\pm$0.12 & 1 & 17.843$\pm$0.103 & 15.882$\pm$0.058 & 18.442$\pm$0.164 & 15.962$\pm$0.025 & 1 & (Y0)       & 1 \\
    CWISEP J094005.50+523359.2   & 09 40 05.46 & +52 33 58.7   & $>$21.26       & 1 & $>$19.845        & 15.913$\pm$0.051 & 18.520$\pm$0.174 & 15.754$\pm$0.022 & 1 & ($\geq$Y1) & 1 \\
    CWISEP J104756.81+545741.6   & 10 47 56.70 & +54 57 41.7   & $>$19.83       & 1 & $>$19.365        & 16.177$\pm$0.064 & 18.731$\pm$0.166 & 16.257$\pm$0.024 & 1 & Y1         & 2 \\
    WISEA J125721.01+715349.3    & 12 57 20.15 & +71 53 49.2   & $>$19.02       & 3 & 18.706$\pm$0.150 & 16.074$\pm$0.046 & 18.889$\pm$0.162 & 16.158$\pm$0.021 & 3 & ($\geq$Y1) & 3 \\
    CWISEP J135937.65-435226.9   & 13 59 37.64 & $-$43 52 27.0 & $>$19.31       & 1 & 19.152$\pm$0.339 & 16.100$\pm$0.068 & 18.169$\pm$0.083 & 15.920$\pm$0.019 & 1 & (Y0)       & 1 \\
    CWISEP J144606.62-231717.8   & 14 46 06.58 & $-$23 17 19.0 & $>$22.36       & 1 & $>$19.648        & 15.955$\pm$0.072 & 18.905$\pm$0.045 & 15.919$\pm$0.018 & 4 & Y1         & 2 \\
    %WISEA J153429.75-104303.3   & 15 34 29.19 & $-$10 43 19.0 & 24.5$\pm$0.3   & 6 & 18.182$\pm$0.189 & 16.145$\pm$0.084 & 16.691$\pm$0.032 & 15.766$\pm$0.023 & 1 & (esdT/Y)   & 7 \\
    CWISEP J223022.60+254907.5   & 22 30 22.51 & +25 49 07.3   & $>$21.16       & 1 & 18.672$\pm$0.217 & 16.227$\pm$0.075 & 19.119$\pm$0.292 & 16.290$\pm$0.026 & 1 & ($\geq$Y1) & 1 \\
    WISEA J224319.56-145857.3    & 22 43 19.66 & $-$14 59 00.3 & $>$19.91       & 3 & 18.435$\pm$0.208 & 15.399$\pm$0.047 & 17.765$\pm$0.058 & 15.336$\pm$0.018 & 3 & (Y0)       & 3 \\
    CWISEP J235547.99+380438.9   & 23 55 48.08 & +38 04 39.3   & 20.28$\pm$0.10 & 1 & 19.332$\pm$0.356 & 15.936$\pm$0.054 & 18.444$\pm$0.259 & 15.926$\pm$0.026 & 1 & (Y0)       & 1 \\
    CWISEP J235644.78-481456.3   & 23 56 44.87 & $-$48 14 56.7 & 21.77$\pm$0.28 & 1 & 18.843$\pm$0.214 & 16.066$\pm$0.057 & 18.727$\pm$0.218 & 16.040$\pm$0.024 & 1 & (Y0.5)     & 1 \\
    \enddata

    \tablecomments{Coordinates are from the 20pc census compiled by \citet{2024ApJS..271...55K}, except for CWISEP J135937.65-435226.9 for which we get the coordinates from the CatWISE2020 catalog. W1 and W2 magnitudes are from CatWISE2020 \citep{2021ApJS..253....8M}. Spectral types listed between parenthesis are photometric estimates.}
    
    \tablerefs{1 - \citet{2020ApJ...889...74M}; 2 - \citet{2024ApJ...973..107B}; 3 - \citet{2020ApJ...899..123M}; 4 - \citet{2020ApJ...888L..19M}.}

\end{deluxetable*}
\end{rotatetable*}

\section{HST/WFC3 observations} \label{sec:data}
The HST/WFC3 observations were performed using the F110W filter, since Y dwarfs have the highest SNR/FWHM in the F110W band (see e.g. \citealt{2015ApJ...804...92S}), minimizing the required integration time. The exposure times, number of orbits, SNR achieved, and the epoch of our HST/WFC3 observations are listed in Table~\ref{tab:observations}. 

Observations were designed to achieve a parallax precision of 10\% or better. This is motivated by the need to  effectively correct for the Lutz-Kelker effect, a systematic error inherent in the measurements of trigonometric parallax for a volume-limited set of stars \citep{1973PASP...85..573L}. Near the maximum distance, d$_{max}$, the volume of stars just inside, (d$_{max}$ - $\Delta$d), is smaller than that just outside, (d$_{max}$ + $\Delta$d), meaning that there are more stars able to scatter into the volume than can scatter outside. This means that the true average parallax is smaller than the average parallax that is measured. The astrometric error needed to correct this effect must be $<$ 17.5\% or else the effect is uncorrectable (see Table 1 in \citealt{1973PASP...85..573L}). Limiting the astrometric error to 10\% reduces the absolute magnitude correction for this effect to 0.11 mag.

The robust identification of unresolved binaries also drove our precision requirement. In the case of an equal-mass (equal-luminosity) binary, the unresolved pair is 0.75 mag brighter than single objects of the same temperature. A parallax error of 17.5\% corresponds to an uncertainty on the absolute magnitude of $\sim$0.4 mag. Therefore, an unresolved binary would be $<$ 2 $\sigma$ overluminous. If the parallax error is brought down to 10\%, the uncertainty on the absolute magnitude is reduced to $\sim$0.2 mag, so overluminous objects can be flagged with $>$ 3 $\sigma$ confidence. 

As discussed in Section~\ref{sec:selection}, for all of the targets earlier WISE/NEOWISE and \textit{Spitzer} epochs are publicly available, and allow us to measure the proper motion and disentangle it from the parallax. The WISE/NEOWISE data cover a long time baseline ($\sim$10 years) and is crucial to securely measure the targets' proper motion, but it is of insufficient precision to detect parallax motion on our faint targets (see e.g. \citealt{2019ApJ...881...17M}). The \textit{Spitzer} data on the other hand are of much greater precision, but for most targets provide too sparse a sampling of the parallactic ellipse to achieve the 10\% goal. In some cases, the \textit{Spitzer} data samples very well one vertex of the parallactic ellipse, but provides only one measurement at the opposite vertex, limiting the precision and reliability of the results. 

In particular, we divided the targets into two categories: (1) targets with \textit{Spitzer} observations at each vertex of the parallactic ellipse; (2) targets with \textit{Spitzer} observations at only one vertex of the parallactic ellipse. For targets pertaining to category (1) \citet{2021ApJS..253....7K} measured a preliminary parallax, whose uncertainty is however too large ($>$17.5\%) to correct for the Lutz-Kelker bias or identify unresolved binaries. For targets pertaining to category (2) a parallax measurement was impossible. To estimate the number of HST observations needed to reach our desired precision, we generated synthetic HST positions, and ran those through the parallax-measuring pipeline of \citet{2019ApJS..240...19K,2021ApJS..253....7K}, combining them with the previous WISE/NEOWISE and \textit{Spitzer} data. The synthetic HST positions were generated by propagating the target's coordinates from a fiducial epoch to later epochs randomly chosen within a 7-day window centered around the time of maximum parallax factor (i.e. the time when the target is at one vertex of the parallactic ellipse). While the ideal measurements would be taken at exactly the maximum parallax factor, this would have put excessive pressure on the scheduling. \citet{2019ApJS..240...19K} found that relaxing timing constraints to within $\sim$3.5 days of maximum parallax factor does not impact the accuracy and precision of the measured parallax and simultaneously increases the schedulability of the observations. The propagation from the fiducial epoch to this 7-day window was done assuming the target has proper motion equal to the value in \citet{2021ApJS..253....7K}, while for the parallax we used the \citet{2021ApJS..253....7K} value when available. Otherwise, we assumed the photometric parallax. Finally, we added gaussian noise to the synthetic positions with $\sigma$ = 3 mas, because \citet{2020MNRAS.494.2068B} found that the residuals between \textit{Gaia} DR2 and HST positions for a set of reference stars are always better than $\sim$3 mas once the distortion correction provided by HST is applied. While our data reduction technique (described in Section~\ref{sec:data}) differs from the one presented in \citet{2020MNRAS.494.2068B} and, therefore, their results are not directly applicable to our targets, the 3\,mas precision floor was a useful approximation for observation planning. For a per-epoch-precision of 3 mas and a desired parallax error of $<$10\%, we found that for targets in category 1 one observation at the vertex of the parallactic ellipse that was previously poorly sampled with \textit{Spitzer} was sufficient. For objects pertaining to category 2, we needed one observation at each vertex of the parallactic ellipse.

\begin{deluxetable*}{l c c c c}
    \tablehead{\colhead{Target ID} & \colhead{Exp. Time} & \colhead{N. orbits} & \colhead{SNR} & \colhead{UT Epoch} \\ 
    \colhead{} & \colhead{s} & \colhead{} & \colhead{} & \colhead{YYYY-MM-DD} }

    \tablecaption{Summary of our HST/WFC3 observations \label{tab:observations}}

    \startdata
    CWISEP J023842.60$-$133210.7 &  2397 & 1 & 9  & 2021-01-17 \\
    CWISEP J040235.55$-$265145.4 &  2397 & 1 & 4  & 2021-02-18 \\
    CWISEP J085938.95+534908.7   &  2397 & 1 & 13 & 2020-11-01 \\
    CWISEP J093852.89+063440.6   &  2397 & 1 & 13 & 2021-02-10 \\
    CWISEP J094005.50+523359.2   &  2397 & 1 & 10 & 2020-11-13 \\
    CWISEP J104756.81+545741.6   &  2397 & 1 & 15 & 2020-11-29 \\
    WISEA J125721.01+715349.3    &  7191 & 3 & 5  & 2021-01-02 \\
    WISEA J125721.01+715349.3    &  7191 & 3 & 4  & 2021-09-13 \\
    CWISEP J135937.65$-$435226.9 &  2397 & 1 & 18 & 2021-01-21 \\
    CWISEP J144606.62$-$231717.8 & 11985 & 5 & 6  & 2021-02-02 \\
    CWISEP J223022.60+254907.5   &  9588 & 4 & 6  & 2020-11-24 \\
    WISEA J224319.56$-$145857.3  &  2397 & 1 & 17 & 2020-11-27 \\
    WISEA J224319.56$-$145857.3  &  2397 & 1 & 16 & 2021-05-29 \\
    CWISEP J235547.99+380438.9   &  2397 & 1 & 21 & 2020-12-19 \\
    CWISEP J235547.99+380438.9   &  2397 & 1 & 21 & 2021-09-06 \\
    CWISEP J235644.78$-$481456.3 &  2397 & 1 & 10 & 2020-12-22 \\
    CWISEP J235644.78$-$481456.3 &  2397 & 1 & 7  & 2022-06-22 \\
    \enddata
\end{deluxetable*}

\section{Data reduction} \label{sec:astrometry}
The HST/WFC3 images were reduced by the HST pipeline\footnote{\url{https://www.stsci.edu/hst/instrumentation/wfc3/software-tools}}. We downloaded the calibrated, distortion-corrected, ``drizzled'' images from MAST\footnote{The specific HST observations used in this paper can be accessed via \dataset[10.17909/bb60-k406]{http://dx.doi.org/10.17909/bb60-k406}},
and refined their astrometric calibration by finding \textit{Gaia} sources in the HST images and fitting a transformation between their HST and \textit{Gaia} DR3 coordinates. 

First, we detected and measured the centroid for all sources in the drizzled images using \textit{imcore}\footnote{\url{http://casu.ast.cam.ac.uk/surveys-projects/software-release/imcore}}. This program implements the methodology described in \citet{1985MNRAS.214..575I, 1997ilt..book.....R}, and here we summarize the key steps. \textit{imcore} first estimates a coarse grid of background values by partitioning the input image in segments of 64$\times$64 pixels. The background values are estimated with a robust iterative k-sigma clipped median. Local background for each pixel is determined using bilinear interpolation of the background grid. After removing this local background, a global sky level and noise level are estimated using the same robust iterative k-sigma clipped median. Source detection is performed by applying a Gaussian filter to the image and searching for groups of pixels above a given threshold and above a specified minimum size. In our analysis, we used a FWHM of 1.25 pixels for the Gaussian filter, a threshold of 4 (defined as number of sigmas above sky level) and a minimum size of 6 pixels. \textit{imcore} then uses a maximum-likelihood method to determine the parameters (and their uncertainties) of an elliptical gaussian distribution for each detected source \citep[Equations 22 and 23 in][]{1985MNRAS.214..575I}. The centroid of each source is measured using an intensity-weighted center-of-gravity method \citep[Equations 28 and 29 in][]{1985MNRAS.214..575I}. 

The x, y pixel coordinates measured by \textit{imcore} are converted to $\alpha, \delta$ using the drizzled image WCS, and then matched to \textit{Gaia} DR3 using a 3'' matching radius. The positions of matching \textit{Gaia} sources are propagated from the \textit{Gaia} epoch (2016.0) to the epoch of the HST observations ($\sim2021$) using their measured parallax and proper motion. We then derived a transformation between the HST and the \textit{Gaia} coordinates by projecting both coordinates onto a tangent plane whose tangent point is defined by the CRVAL1 and CRVAL2 FITS header keywords. We then solved for the parameters of the transformation using the IDL routine \textit{mpfit} \citep{2009ASPC..411..251M}, which implements the Levenberg–Marquardt least-squares algorithm. The residuals of the transformation are added in quadrature to the measurement errors to compute the final coordinate uncertainties. This assumes that the \textit{Gaia} propagated positions are the true positions for the reference stars. However, given that the vast majority of our reference stars are faint (\textit{Gaia} G $>$ 19 mag), additional care must be taken since \textit{Gaia} proper motions in this brightness regime can be poorly determined \citep[see, e.g.,][]{2021jwst.rept.7716A}. To validate our methodology, we repeated the astrometric registration without propagating the \textit{Gaia} coordinates to the epoch of observation, and compared the transformation residuals with those obtained using propagated coordinates. The median residuals obtained with propagated coordinates are $\sim$5\,mas, while those obtained without propagating the \textit{Gaia} coordinates are $\sim$30\,mas. For all fields, the residuals obtained propagating the \textit{Gaia} coordinates are better then those obtained without.

The number of parameters for the transformation depends on the number of reference stars available. Ideally, we would like to fit for a six-parameter transformation, which accounts for offsets, rotation, skew, and scaling. However, in many fields we have only a handful or fewer reference stars, forcing us to use simpler transformations. If the number of reference stars is $>10$, we fit for the full six-parameter transformation; if the number of reference stars is between 3 and 10, we fit for a three-parameter transformation, which accounts only for offset and rotation. Finally, if fewer than 3 reference stars are available, we only compute the median offset, along each axis, between \textit{Gaia} and HST (equivalent to fitting for a two-parameter transformation).

We then combined the HST measurement with the \textit{Spitzer} observations presented in \citet{2021ApJS..253....7K}\footnote{The \textit{Spitzer} observations used in this paper can be acessed via \dataset[doi:10.26131/IRSA543]{https://catcopy.ipac.caltech.edu/dois/doi.php?id=10.26131/IRSA543})} and the unWISE measurements from \citet{2023AJ....165...36M}, and fit them with a 5-parameter astrometric model, following the procedure described in detail in \citet[their Section 5.2.3]{2019ApJS..240...19K}, summarized here. Let us call our measured positions $\alpha(t), \delta(t)$. The 5-parameter model is defined so that:
\begin{eqnarray}
    \label{eq:alphares}
    \alpha_{residual} & = & (\alpha(t) - \alpha')/\cos{\delta'} \\
    & & - \varpi (X(t)\sin{\alpha'} -Y(t)\cos{\alpha'})/3600 \nonumber 
\end{eqnarray}
\begin{eqnarray}
    \label{eq:deltares}
    \delta_{residual} & = & \delta(t) - \delta' - \varpi (X(t)\cos{\alpha'}\sin{\delta'} \\
    & & + Y(t)\sin{\alpha'}\sin{\delta'} - Z(t)\cos{\delta'})/3600 \nonumber
\end{eqnarray}
\noindent where
\begin{equation}
    \alpha' = \alpha_0 + (\mu_\alpha (t-t_0)/\cos{\delta'})/3600
\end{equation}
\begin{equation}
    \delta' = \delta_0 + \mu_\delta (t-t_0) / 3600
\end{equation}

The reference epoch, $t_0$, is chosen to be the mid-point of the time span covered by the full set of observations (WISE/NEOWISE + \textit{Spitzer} + HST), while $\alpha_0, \delta_0, \mu_\alpha, \mu_\delta,$ and $\varpi$ are all parameters of the fit. The coordinates of the observatory at each epoch ($X(t), Y(t), Z(t)$) are known. The residuals defined in Equations~\ref{eq:alphares} and \ref{eq:deltares} are minimized using the aforementioned \textit{mpfit}, which returns the best-fit parameters and their uncertainties (computed from the covariance matrix).

\begin{figure}
    \centering
    \includegraphics[width=0.9\linewidth]{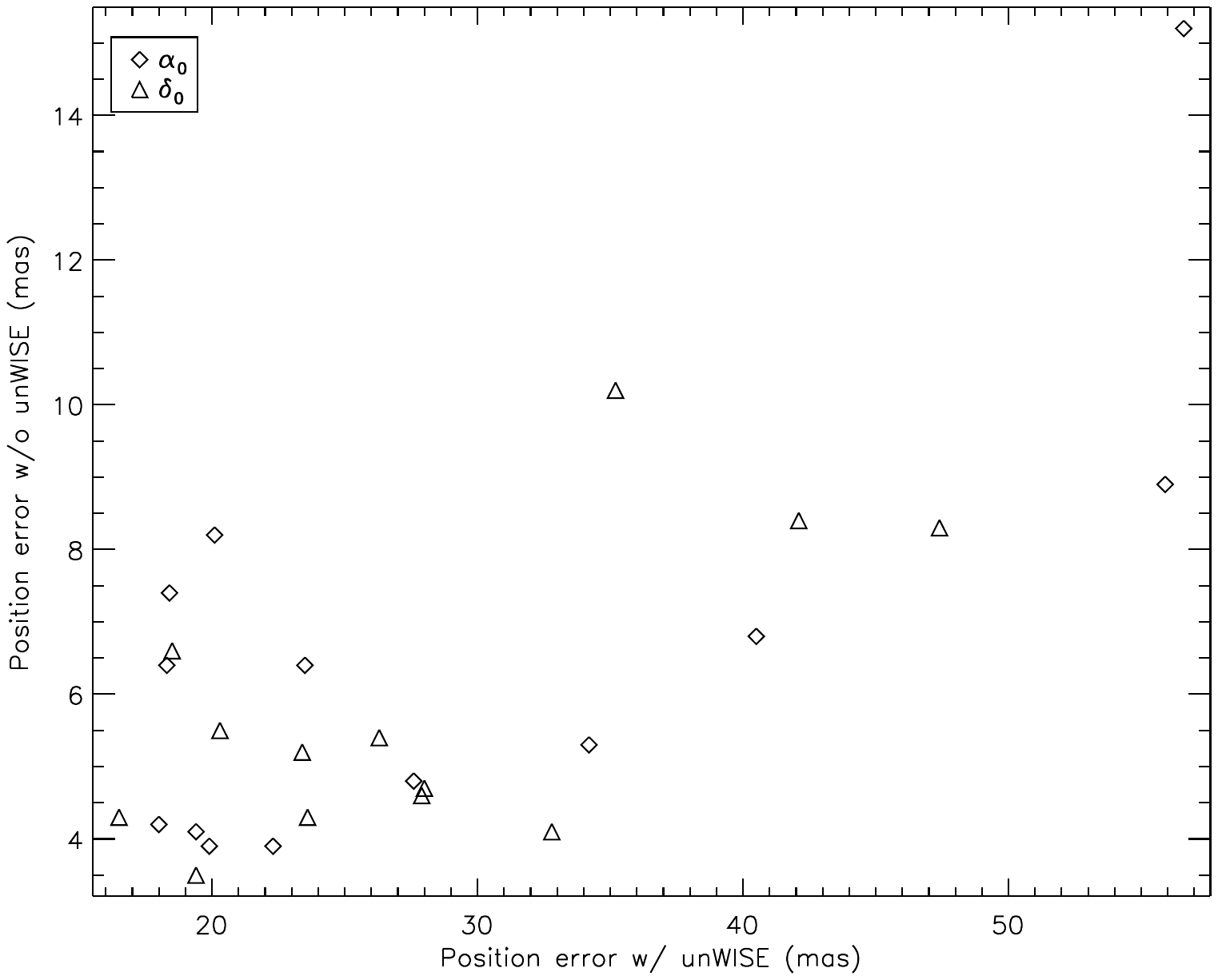}
    \includegraphics[width=0.9\linewidth]{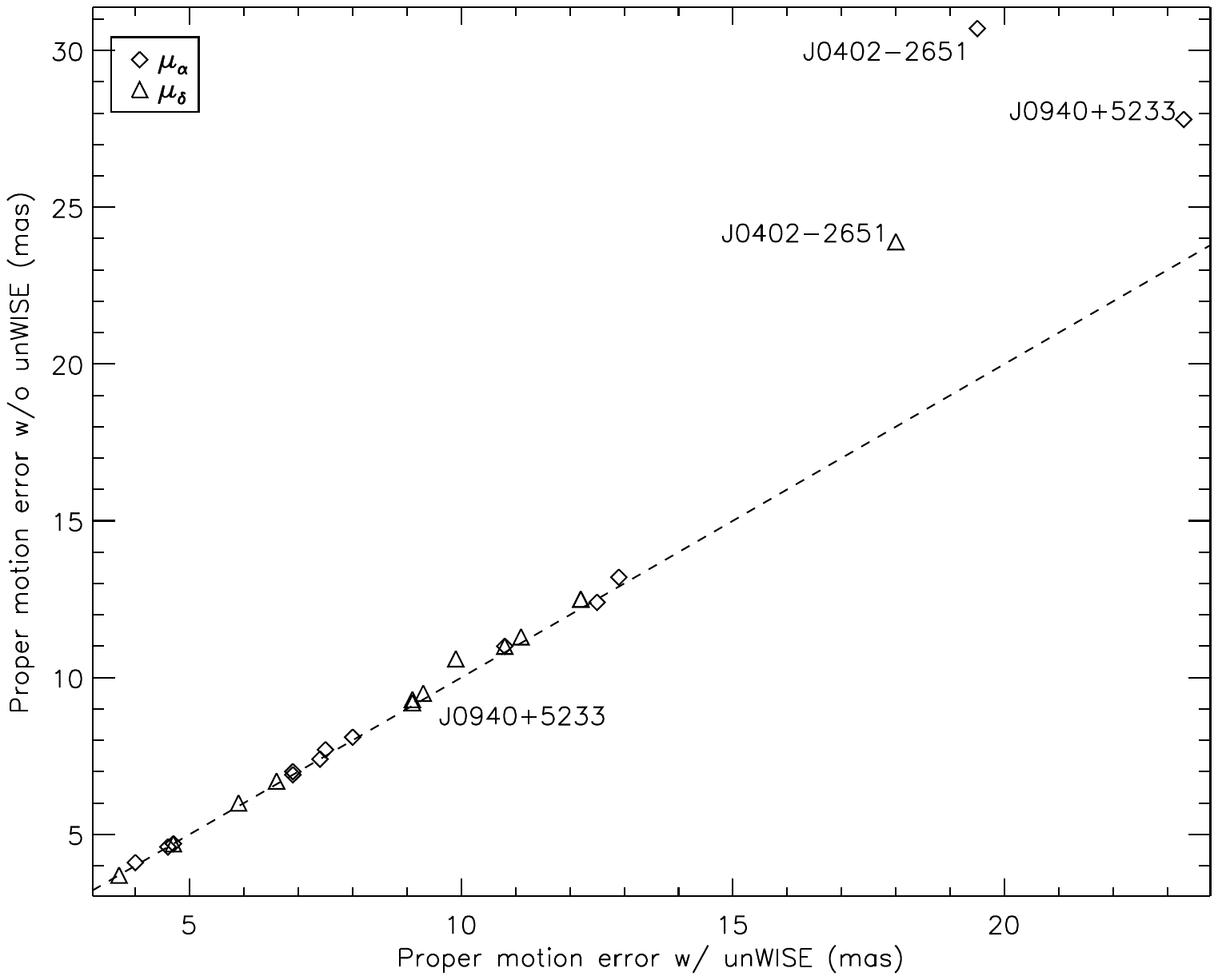}
    \includegraphics[width=0.9\linewidth]{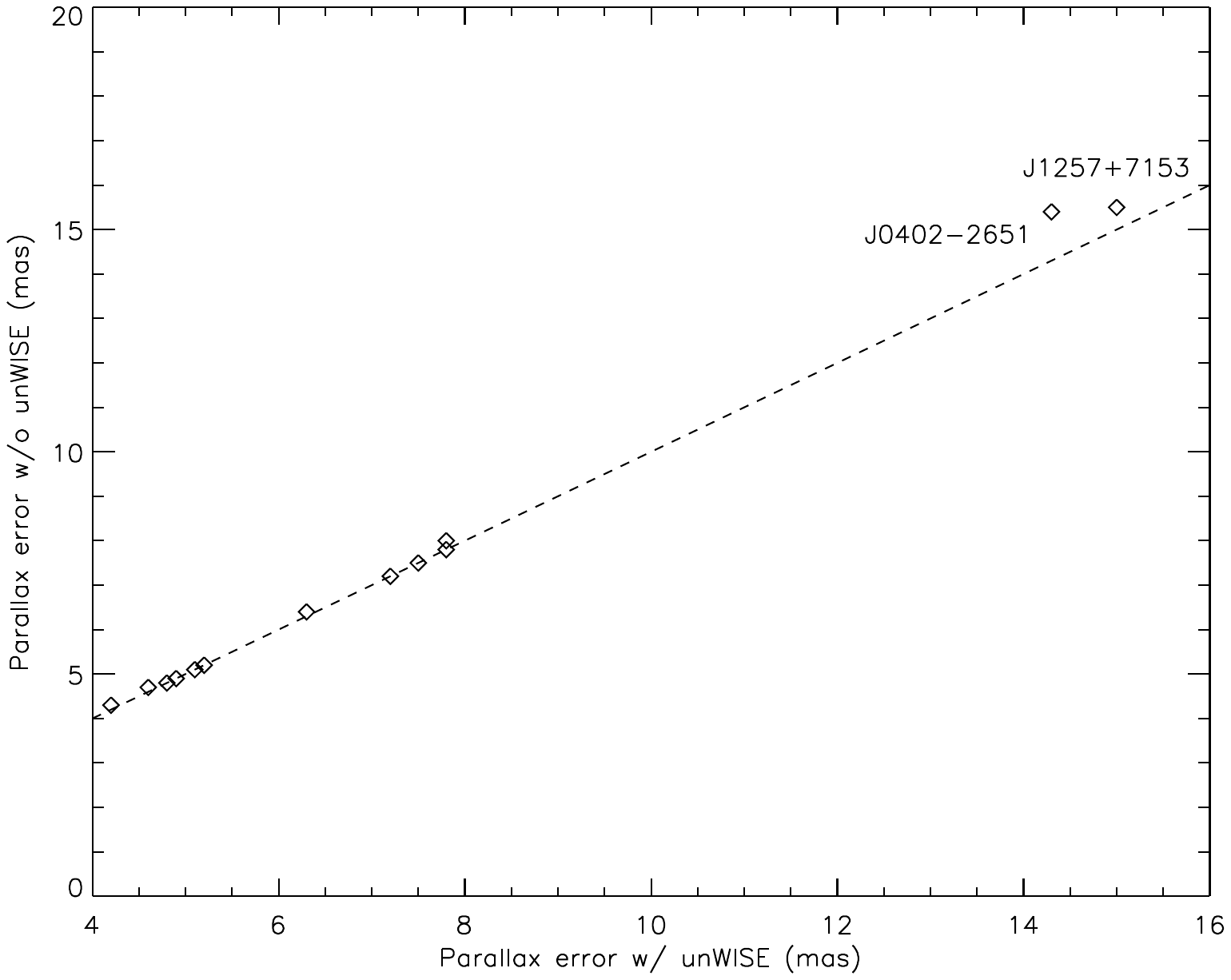}
    \caption{Comparison of the uncertainties on the best-fit parameters of our astrometric model obtained when using the full unWISE + Spitzer + HST dataset vs. using only the Spitzer + HST data. Uncertainties on the reference coordinates (top panel) are greatly improved when omitting unWISE data, while those on proper motion (middle panel) and parallax (bottom panel) are mostly unchanged. Objects with more uncertain proper motions (i.e. $\sigma_\mu > 15$ mas yr$^{-1}$) or parallaxes (i.e. $\sigma_\varpi > 10$ mas) show the largest deviations from the identity line (dashed line).}
    \label{fig:astro_comparison}
\end{figure}

Following advice from the referee, we also fit the 5-parameter astrometric model to the \textit{Spitzer} and HST data only (omitting the unWISE data). Despite the much shorter baseline covered by the \textit{Spitzer} and HST data alone ($\sim$2 years vs. $\sim$10 years when including unWISE data), the new fit yielded similar uncertainties on proper motions and parallaxes, and significantly smaller uncertainties on the $\alpha_0, \delta_0$ coordinates at the reference epochs. The best-fit values are consistent between the two fit well within the uncertainties. Figure~\ref{fig:astro_comparison} shows the uncertainties on coordinates (top panel), proper motions (middle panel) and parallaxes (bottom panel) for the fit with and without unWISE data. The typical $\alpha_0,\delta_0$ uncertainties decrease from the 15--60 mas range to the 3--16 mas range when omitting the unWISE data. For objects with well-measured parallaxes and proper motions, i.e. $\sigma_\varpi < 10$\,mas and $\sigma_\mu < 15$\, mas yr$^{-1}$, the corresponding uncertainties remain practically the same. For objects with more uncertain parallaxes or proper motion, i.e. $\sigma_\varpi \geq 10$\,mas or $\sigma_\mu \geq 15$\, mas yr$^{-1}$, the corresponding uncertainties increase noticeably when omitting the unWISE data. In both cases the best-fit parallax and proper motion values are consistent within much less than one $\sigma$. Given the importance of having well-measured reference positions at a given epoch (to be used e.g. for reliable placement of the target in the slit of a spectrograph) we chose to publish the parameters obtained when omitting the unWISE data.

\begin{figure*}
    \centering
    \includegraphics[width=\linewidth, trim={0cm 0cm 0cm 0.8cm}, clip]{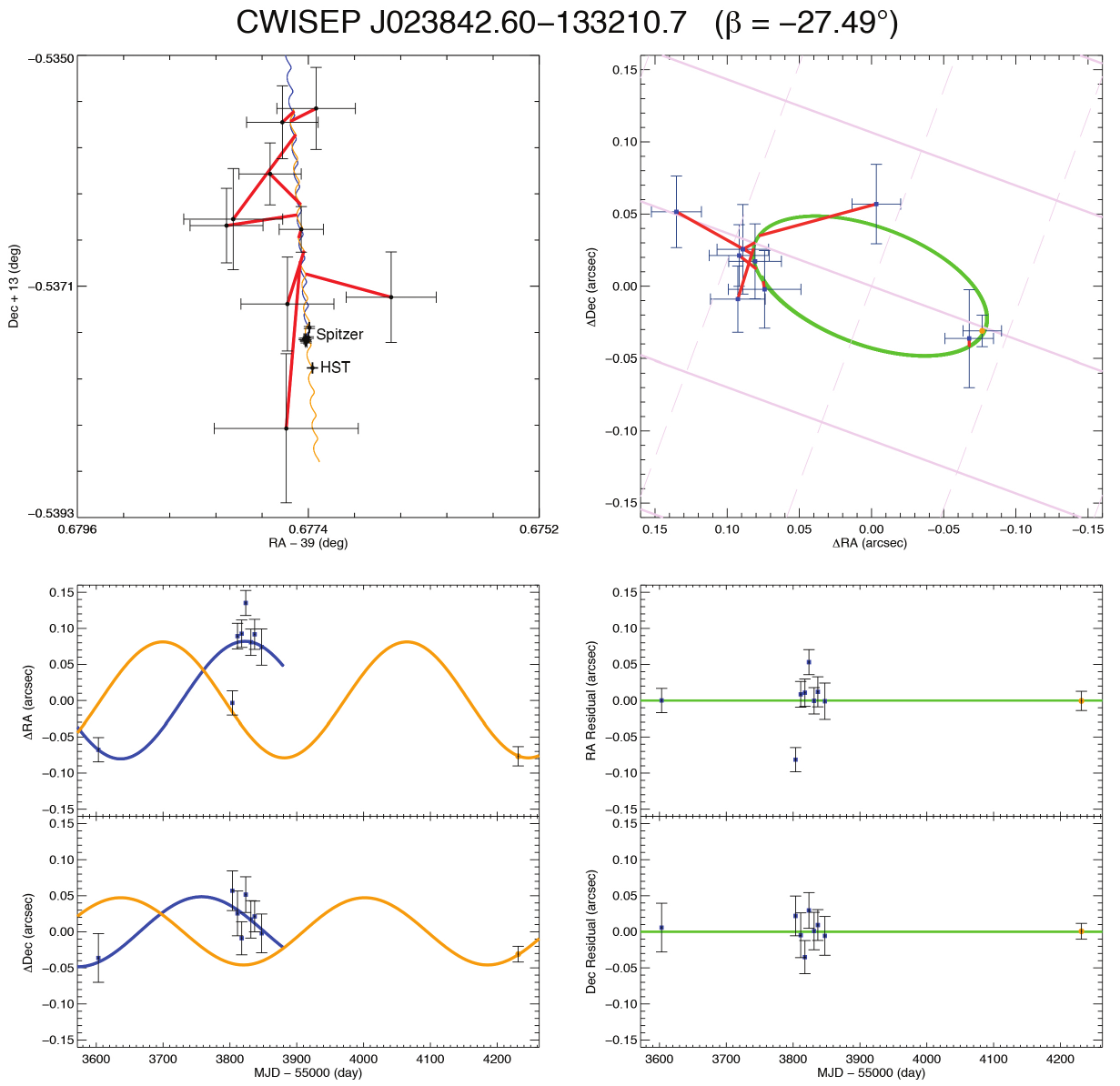}
    \caption{Astrometric fit for CWISEP J023842.60$-$133210.7. (Upper left) A square patch of sky showing the measured coordinates and their uncertainties at each epoch (black points with error bars). Points with small error bars are the \textit{Spitzer} and HST  measurements, labeled for clarity; those with larger error bars are the unWISE measurements. The blue curve shows the best fit from the vantage point of \textit{Spitzer}. The orange curve shows the same fit as seen from the vantage point of WISE/NEOWISE and HST (i.e. the Earth). Red lines connect each observation to its corresponding point along the best-fit curve. (Upper right) A square patch of sky centered at the mean position of the target. The green ellipse is the parallactic fit. For clarity, only the \textit{Spitzer} and HST measurements are shown, in blue and orange respectively. Though not shown, the unWISE measurements are included in the fit. Again, red lines connect the time of the observation with its prediction. In the background is the ecliptic coordinate grid, with lines of constant $\beta$ shown in solid pale purple and lines of constant $\lambda$ shown in dashed pale purple. (Lower left) The change in R.A. and decl. as a function of time with the proper motion component removed. The parallactic fit from the vantage point of \textit{Spitzer} is shown in blue, and from the vantage point of HST is shown in orange. Again, only the \textit{Spitzer} and HST measurements are shown, in blue and orange respectively. (Lower right) The R.A. and decl. residuals from the fit as a function of time. As with the lower left panel, only the \textit{Spitzer} and HST data are shown, in blue and orange respectively.}
    \label{fig:plx_fit_unwise}
\end{figure*}

\begin{figure*}
    \centering
    \includegraphics[width=\linewidth, trim={0cm 4cm 0cm 4.5cm}, clip]{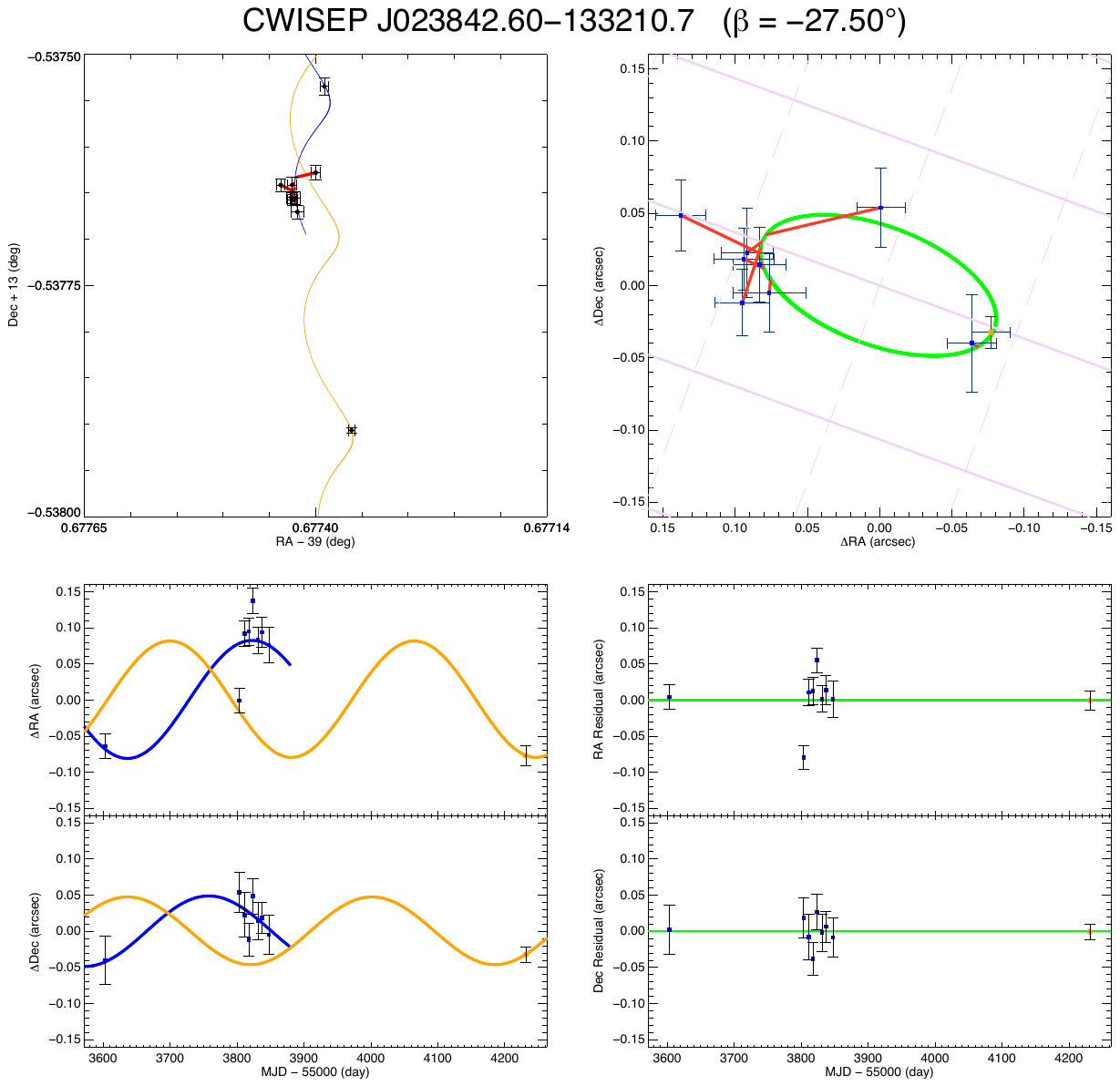}
    \caption{Same as Figure~\ref{fig:plx_fit_unwise}, but without using unWISE data.}
    \label{fig:plx_fit_no_unwise}
\end{figure*}

The results are presented in Table~\ref{tab:astrometry}, while in Figure~\ref{fig:plx_fit_unwise} and \ref{fig:plx_fit_no_unwise} we show an example astrometric fit. For all but one of our targets (WISEA J125721.01+715349.3) we achieved uncertainties $<$17.5\% on the parallax measurement, which is the requirement to effectively correct for the Lutz-Kelker effect (see Section \ref{sec:data}). WISEA J125721.01+715349.3 is much fainter in the F110W band than we had estimated using its WISE/NEOWISE and \textit{Spitzer} photometry. This resulted in low signal-to-noise ratio (S/N) detections at both HST epochs, which in turn resulted in larger uncertainties on the measured positions, preventing us from obtaining a more precise parallax measurement. 

For several of our targets we did not reach our target precision of $<$10\% uncertainty on the parallax measurement. This is partly due to underestimation of the F110W magnitudes leading to low S/N detections, and partly due to the lack of \textit{Gaia} stars in the field of view preventing us from performing a more robust astrometric registration of the HST images. WFC3 has a relatively narrow field of view of $\sim$2$\times$2 arcmin. In most cases this yielded only between 2 and 6 reference stars per image, forcing us to use a simpler 3-parameter re-registration, but also forcing us to use all reference stars without applying any quality cut on their astrometry (e.g. removing stars with \textit{ruwe}$>$1.4). This resulted in large residuals after the re-registration fit, which in turn resulted in larger positional uncertainties for our targets.  

\begin{deluxetable*}{l c c c c c c c c c c c c c}
    \tablehead{\colhead{Target ID} & \colhead{$\alpha_0$} & \colhead{$\sigma_\alpha$} & \colhead{$\delta_0$} & \colhead{$\sigma_\delta$} & \colhead{$t_0$} & \colhead{$\varpi$} & \colhead{$\mu_\alpha$} & \colhead{$\mu_\delta$} & \colhead{\textit{nref}} & \colhead{\textit{npar}} & \colhead{\textit{nep}} & \colhead{$\Delta t$} & \colhead{$\chi^2$/dof} \\ 
    \colhead{} & \colhead{deg} & \colhead{mas} & \colhead{deg} & \colhead{mas} & \colhead{MJD} & \colhead{mas} & \colhead{mas yr$^{-1}$} & \colhead{mas yr$^{-1}$} & \colhead{} & \colhead{} & \colhead{} & \colhead{yr} & \colhead{} }

    \tablecaption{Our astrometric results \label{tab:astrometry}}

    \startdata
    CWISEP J023842.60$-$133210.7 &  39.677397 &  6.8 & -13.537668 &  8.4 & 58845.23 &  84.6$\pm$7.5  &  -53.6$\pm$12.4 & -782.3$\pm$12.5 & 4  & 3 & 9 & 1.720 & 2.940 \\
    CWISEP J040235.55$-$265145.4 &  60.649781 & 15.2 & -26.863713 &  8.3 & 58866.67 & 104.5$\pm$15.4 &  777.1$\pm$30.7 & -519.3$\pm$23.9 & 2  & 2 & 9 & 1.808 & 0.402 \\
    CWISEP J085938.95+534908.7 &   134.911683 &  3.9 &  53.818626 &  5.2 & 58740.94 &  57.9$\pm$4.7  & -161.6$\pm$7.7  & -316.0$\pm$9.2  & 3  & 3 & 7 & 1.683 & 0.479 \\
    CWISEP J093852.89+063440.6 &   144.721207 &  6.4 &   6.576888 &  4.3 & 58779.48 &  65.3$\pm$7.2  &  449.6$\pm$8.1  & -626.4$\pm$9.3  & 5  & 3 & 7 & 1.916 & 1.599 \\
    CWISEP J094005.50+523359.2 &   145.022152 &  8.9 &  52.565964 &  5.4 & 58748.02 &  68.4$\pm$8.0  & -321.7$\pm$27.8 & -332.1$\pm$9.5  & 2  & 2 & 7 & 1.708 & 1.415 \\
    CWISEP J104756.81+545741.6 &   161.985310 &  4.1 &  54.961352 &  4.7 & 58757.73 &  68.3$\pm$4.9  & -510.4$\pm$6.9  &  -93.8$\pm$11.3 & 5  & 3 & 7 & 1.754 & 3.608 \\
    WISEA J125721.01+715349.3 &    194.329962 &  8.2 &  71.897008 & 10.2 & 58893.63 &  67.7$\pm$15.5 & -937.0$\pm$7.0  &   97.0$\pm$10.6 & 4  & 3 & 5 & 3.087 & 0.413 \\
    CWISEP J135937.65$-$435226.9 & 209.906191 &  3.9 & -43.874511 &  3.5 & 58840.72 &  41.1$\pm$4.3  & -376.8$\pm$7.4  & -193.4$\pm$6.7  & 49 & 6 & 8 & 1.647 & 1.652 \\
    CWISEP J144606.62$-$231717.8 & 221.526155 &  5.3 & -23.289606 &  4.1 & 58845.47 & 103.8$\pm$5.1  & -755.0$\pm$13.2 & -888.8$\pm$12.5 & 12 & 6 & 8 & 1.681 & 1.144 \\
    CWISEP J223022.60+254907.5 &   337.593147 &  4.8 &  25.817978 &  4.6 & 58804.79 &  62.2$\pm$5.2  & -554.4$\pm$11.0 & -434.2$\pm$11.0 & 15 & 6 & 7 & 1.723 & 1.984 \\
    WISEA J224319.56$-$145857.3 &  340.832265 &  6.4 & -14.984111 &  5.5 & 58902.70 &  87.3$\pm$6.4  &  319.6$\pm$4.7  & -547.0$\pm$6.0  & 6  & 3 & 5 & 2.608 & 0.562 \\
    CWISEP J235547.99+380438.9 &   358.951550 &  4.2 &  38.077572 &  4.3 & 58969.05 &  54.4$\pm$4.8  &  707.7$\pm$4.1  &   27.9$\pm$3.7  & 18 & 6 & 4 & 2.851 & 0.570 \\
    CWISEP J235644.78$-$481456.3 & 359.188704 &  7.4 & -48.249257 &  6.6 & 58987.19 &  62.0$\pm$7.8  &  858.4$\pm$4.6  &  -66.8$\pm$4.7  & 4  & 3 & 5 & 3.668 & 1.535 \\
    \enddata

    \tablecomments{$\alpha, \delta$ are the coordinates at epoch MJD. $\mu_\alpha$ includes the $\cos \delta$ factor. \textit{nref} is the number of reference stars used for the astrometric calibration of the HST image, \textit{npar} the number of parameters of the transformation, \textit{nep} the number of epochs, $\Delta t$ the total time span covered by the observations, and $\chi^2$/dof the chi squared of the fit divided by the number of degrees of freedom.}
     
\end{deluxetable*}

\section{Discussion} \label{sec:results}
One target, CWISEP~J144606.62--231717.8, is confirmed to be a member of the 10-pc sample. Another object, CWISEP~J040235.55--265145.4, is tentatively placed within that sample too, but the large uncertainty on its parallax requires additional observations to shore up its distance measurement. CWISEP~J135937.65--435226.9, on the other hand, is now firmly excluded from the 20-pc sample by its well-measured parallax. Four of our 13 targets have absolute \textit{Spitzer} ch2 magnitudes consistent with a spectral type Y0 or later according to the absolute mangnitude to spectral type calibration of \citet{2021ApJS..253....7K} -- CWISEP~J023842.60--133210.7, CWISEP~J040235.55--265145.4, CWISEP~J104756.81+545741.6, and CWISEP~J144606.62--231717.8. \citet{2024ApJ...973..107B} have already spectroscopically confirmed that CWISEP~J104756.81+545741.6 and CWISEP~J144606.62--231717.8 are Y dwarfs.

\begin{figure}
    \centering
    \includegraphics[width=\linewidth, trim={2cm 1cm 3cm 3cm}, clip]{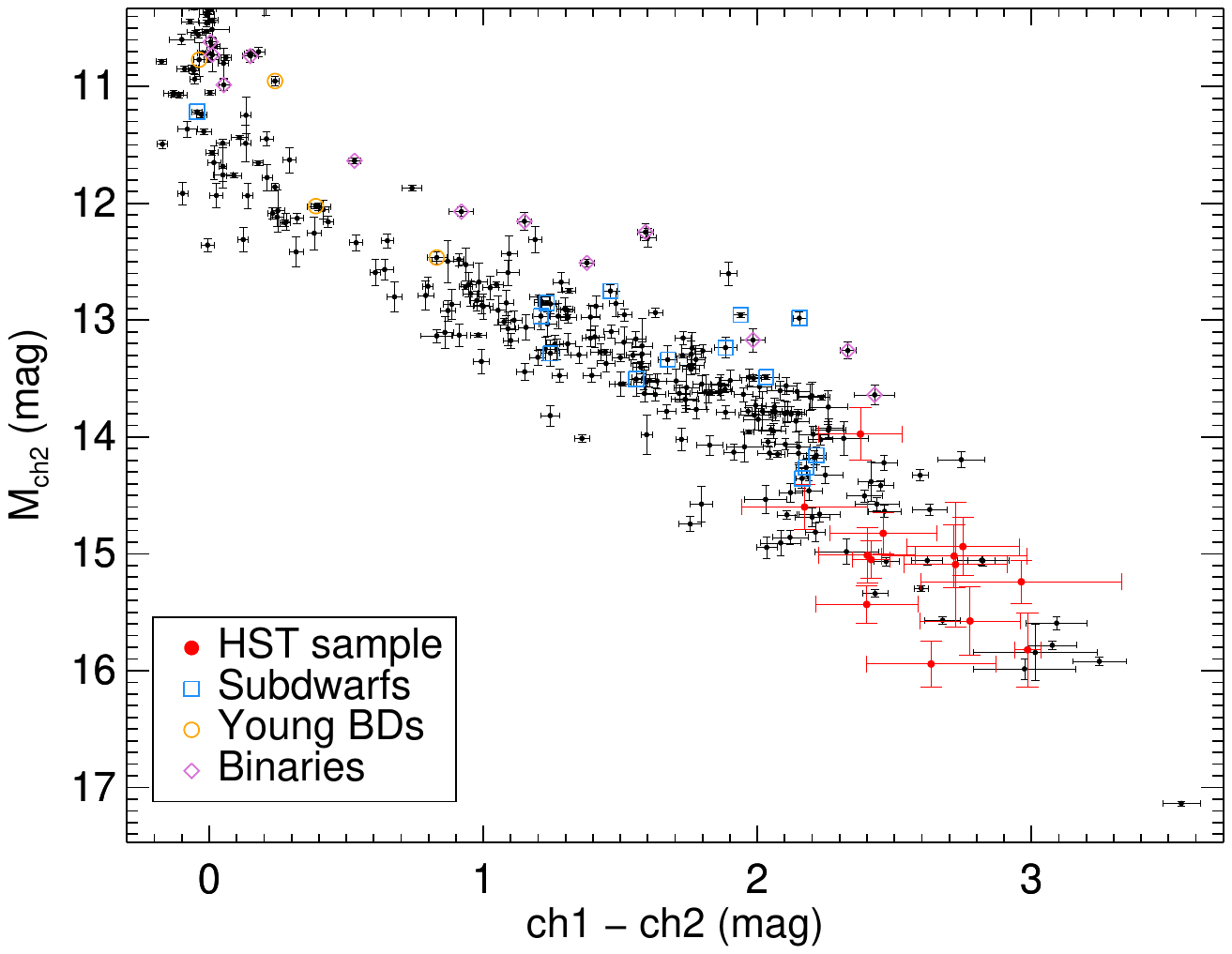}
    \includegraphics[width=\linewidth, trim={2cm 1cm 3cm 3cm}, clip]{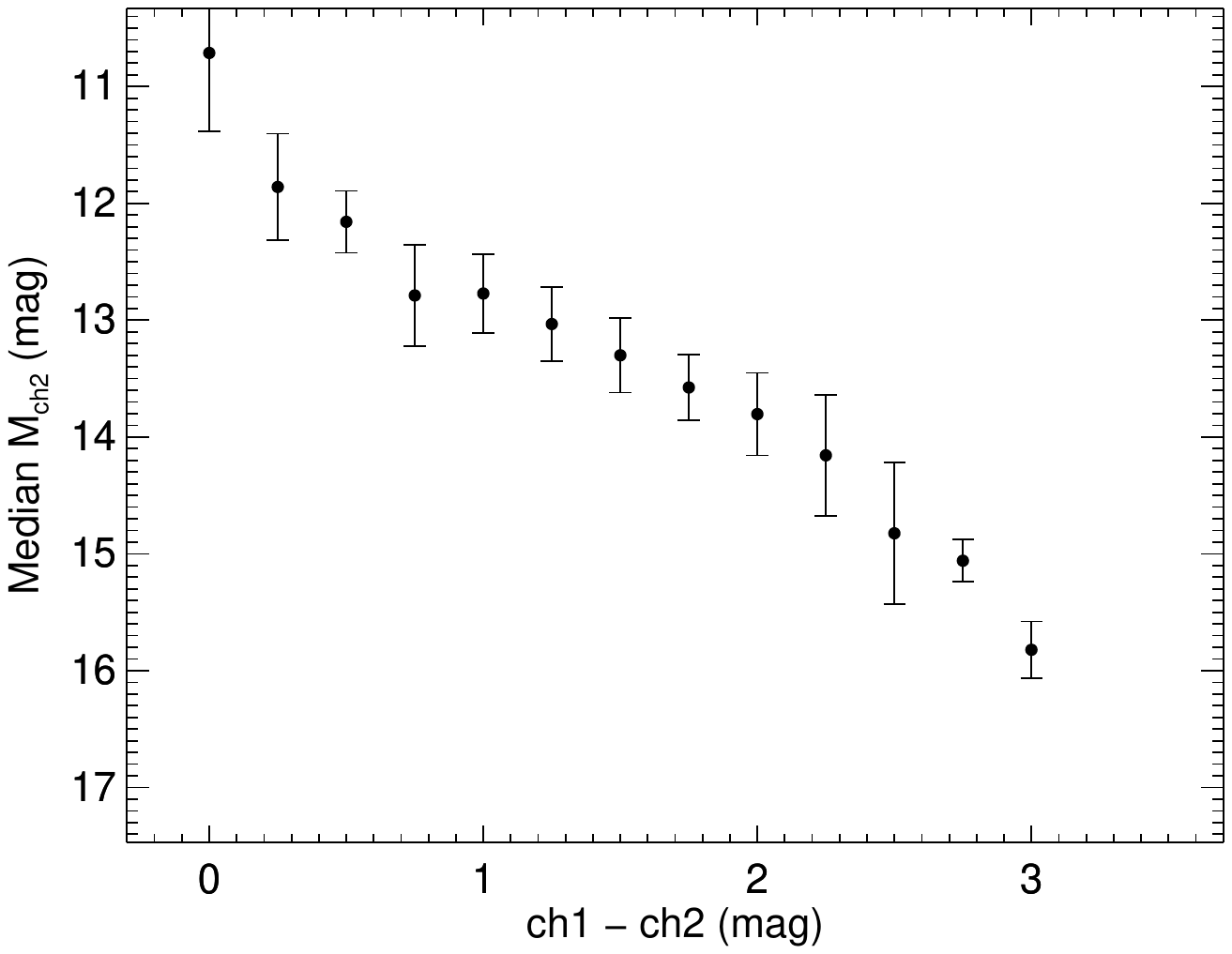}
    \caption{\textit{Top:} Color-magnitude diagram for our sample compared to known nearby L, T, and Y dwarfs from the 20pc census of \citet{2021ApJS..253....7K}. Our targets occupy the bottom of the main sequence, and further highlight the large photometric scatter among the coldest brown dwarfs. \textit{Bottom:} median absolute \textit{Spitzer} ch2 magnitude as a function of ch1--ch2 color, in bins of 0.25 mag. The error bars represent the $1\sigma$ scatter in the same bins.}
    \label{fig:CMD_full}
\end{figure}

\begin{figure}
    \centering
    \includegraphics[width=\linewidth, trim={4cm 1cm 6cm 3cm}, clip]{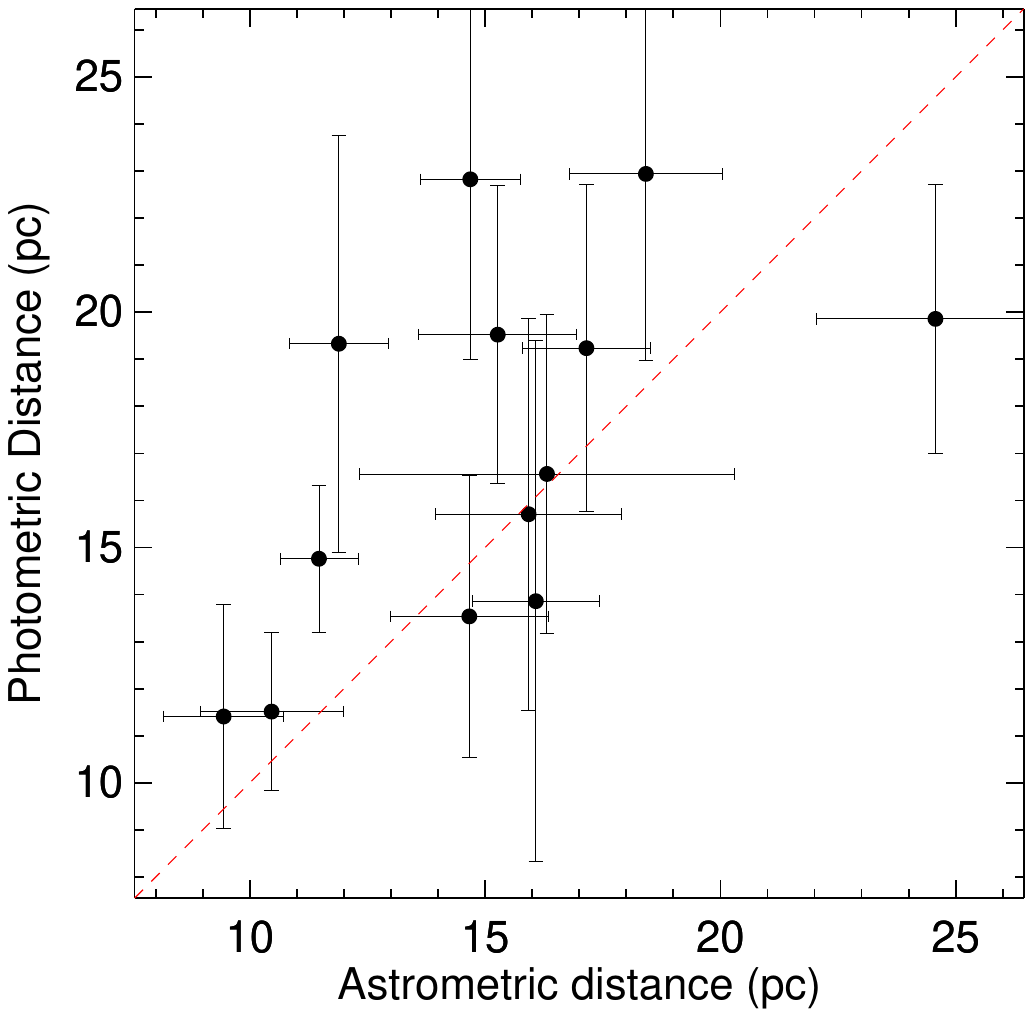}
    \caption{A comparison between our measured astrometric distances and the photometric distances estimated using the \textit{Spitzer} magnitudes and the polynomial relations from \citet{2021ApJS..253....7K}. The large photometric diversity among cold brown dwarfs makes distance estimates often inaccurate and, in general, of much lower precision compared to astrometric measurements. \label{fig:plx_vs_dphot}}
\end{figure}

With the improved astrometry derived here we can also look at the 13 targets of this paper in context with the rest of the nearby substellar population. The top panel of figure~\ref{fig:CMD_full} shows the absolute \textit{Spitzer} ch2 magnitude as a function of the \textit{Spitzer} ch1--ch2 color for our targets, compared to L, T, and Y dwarfs taken from the 20-pc census of \citep{2021ApJS..253....7K}. Our targets are among the coldest and reddest objects in the Solar neighborhood. Objects at the bottom of the main sequence show increased photometric diversity. While warmer, bluer objects (ch1--ch2$\lesssim$1 mag) form a tight sequence, colder, redder objects are spread more widely in color-magnitude space, with the intrinsic scatter of the population becoming particularly noticeable beyond ch1--ch2$\sim$2 mag. The bottom panel of Figure~\ref{fig:CMD_full} shows the median absolute \textit{Spitzer} ch2 magnitude as a function of ch1--ch2 color, computed in bins of 0.25 mag. The error bars are the $1\sigma$ dispersion in the same bins, and show an increase in the population scatter, from $\sim0.3$ mag for sources in the $1 < {\rm ch1-ch2} < 2$ mag range, to $\sim0.5$ mag for sources with $2 < {\rm ch1-ch2} < 2.6$ mag. This trend appears to reverse for even redder objects, however, the two reddest bins are sparsely populated (comprising 8 and 7 objects, respectively) so the true scatter of this population may not have been fully revealed yet. Similar diversity is observed in mid-IR spectra \citep[see e.g.][]{2024ApJ...973..107B} and several factors can contribute to it -- metallicity, surface gravity, disequilibrium chemistry, viewing angle. This intrinsic scatter makes photometric distance estimates highly unreliable for Y dwarfs, so parallax measurements are necessary. This is illustrated in Figure~\ref{fig:plx_vs_dphot}, where we plot our astrometric distances against the photometric distance estimates obtained using the \textit{Spitzer} magnitudes and the polynomial relations from \citet{2021ApJS..253....7K}. The photometric distance estimates are often inconsistent with the measured distances, and for six objects the two distances are discrepant by more than $1\sigma$. In general, even when the two distances are in agreement, the photometric distances have uncertainties that are $\sim2$ or $3\times$ larger than the astrometric distance uncertainties.

\begin{figure}[h]
    \centering
    \includegraphics[width=\linewidth, trim={1.8cm 1cm 3cm 3cm}, clip]{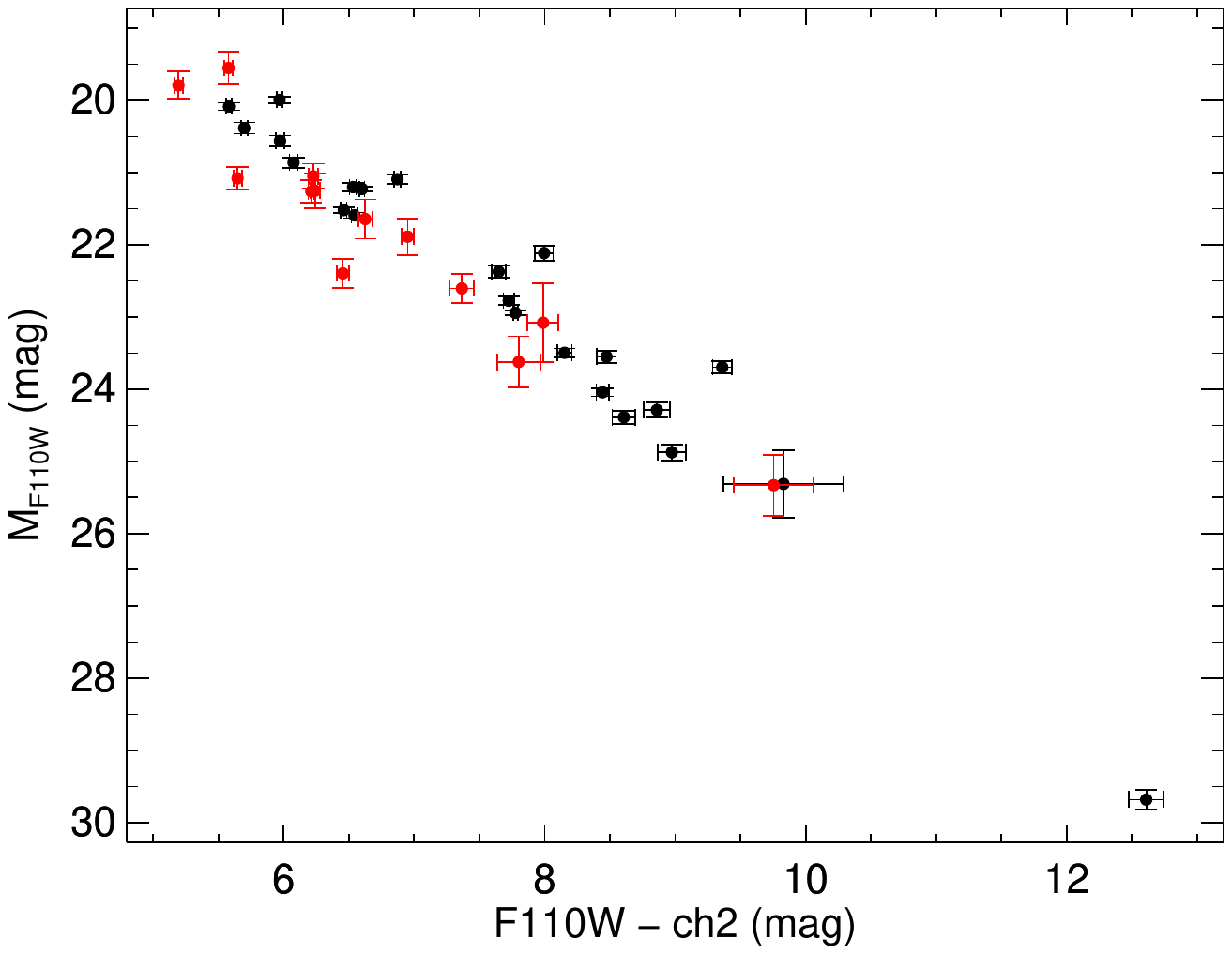}
    \includegraphics[width=\linewidth, trim={1.8cm 1cm 3cm 3cm}, clip]{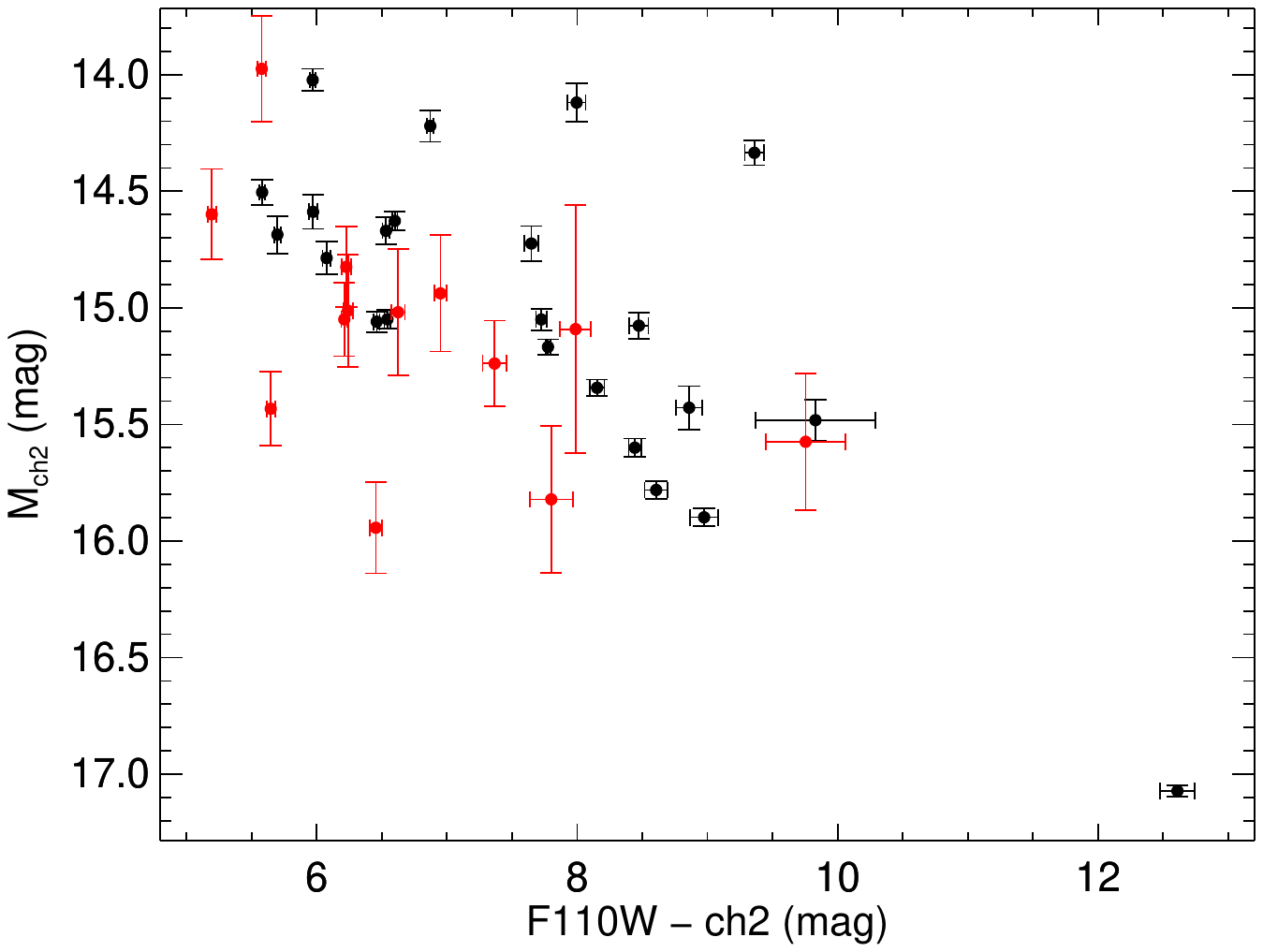}
    \includegraphics[width=\linewidth, trim={1.8cm 1cm 3cm 3cm}, clip]{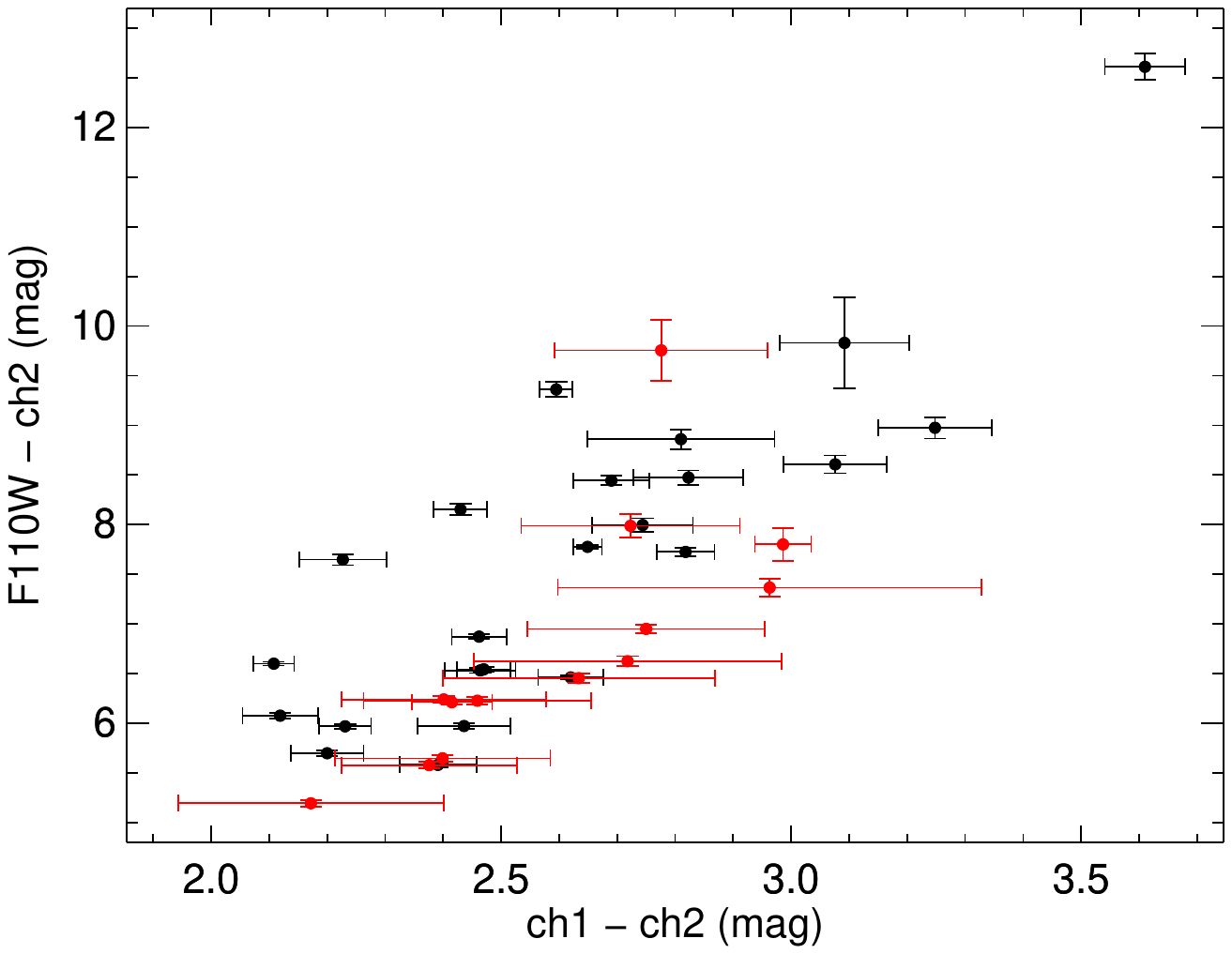}
    \caption{Color-magnitude diagrams for our sample (red points) compared to known nearby T and Y dwarfs (black points) from \citep{2015ApJ...804...92S}. \textit{Top:} Absolute F110W magnitude as a function of F110W--ch2 color. The sample stretches over nearly 10 magnitudes in absolute magnitude, and 7.5 magnitudes in color, clearly showing the rapid collapse of short-wavelength emission in these very cold substellar objects. \textit{Middle:} Absolute ch2 magnitude as a function of F110W--ch2 color. The sequence spans ``only'' $\sim$2 magnitudes in absolute magnitude, but shows larger intrinsic scatter, matching the spectroscopic diversity observed in MIR spectra \citep[see e.g.][]{2024ApJ...973..107B}. \textit{Bottom:} F110W--ch2 color vs. ch1--ch2 color. Objects with similar ch1--ch2 can have F110W--ch2 colors spanning almost 3.5 magnitudes.}
    \label{fig:CMD_HST_only}
\end{figure}

Figure~\ref{fig:CMD_HST_only} compares our sample with nearby cold brown dwarfs with available HST photometry, either measured or derived from HST spectra \citep{2011ApJ...743...50C,2014AJ....147..113C,2021ApJ...920...20C,2012ApJ...753..156K,2016AJ....152...78L,2015ApJ...804...92S,2016ApJ...823L..35S}. The top panel highlights the rapid collapse of the near-infrared portion of cold Y dwarfs SEDs, with the absolute F110W magnitude spanning nearly 10 magnitude in range from the warmer late-Ts to the coldest brown dwarf known \citep[WISE~J0855;][]{2014ApJ...786L..18L}, a range of only $\sim$250\,K. Longer-wavelength flux (middle panel) decreases comparatively more slowly, with the full sample spanning approximately 3 magnitudes over the same range of temperatures. The bottom panel of Figure~\ref{fig:CMD_HST_only} shows the F110W--ch2 color as a function of the ch1--ch2 color. While the two colors are clearly correlated, objects with ch1--ch2 colors within a few tenths of a magnitude of each other can have F110W--ch2 colors spanning almost 3.5 magnitudes. Interestingly, our sample appears to be overall bluer in F110W--ch2 than other Y dwarfs. Older, metal poor objects typically display bluer NIR colors compared to field-age, solar metallicity objects, but the large uncertainties on our \textit{Spitzer} colors demand caution before drawing any firm conclusion on the nature of our targets. 

Figure~\ref{fig:CMD_full} and \ref{fig:CMD_HST_only} highlight how, despite several focused search-and-follow-up campaigns using WISE/NEOWISE data and, more recently, JWST data, there remains a clear gap in color and absolute magnitude between WISE~J0855 and the rest of the cold brown dwarf population. Given the brightness and proximity of WISE~J0855 (W2 = 13.820$\pm$0.029, d $\sim$2.3 pc) and the depth achieved by WISE/NEOWISE, \citet{2014AJ....148...82W} estimated that between 4 and 35 similarly cold brown dwarfs should have been detected. Yet, more than a decade after that estimate, WISE~J0855-like objects remain elusive. 

\section{Conclusions} \label{sec:conclusions}

We presented improved astrometry for 13 of the coldest T and Y dwarfs in the Solar neighborhood. The new astrometry was determined by combining existing \textit{Spitzer} data with our own \textit{HST} observations. By using \textit{Gaia} to re-register all images to a common reference frame, we were able to seamlessly combine the two datasets. Parallax measurements are resource- and time-consuming, since many observations spanning several years are needed to achieve reliable measurements and to disentangle parallactic motion and proper motion. With this approach, we were able to achieve the desired astrometric precision by leveraging existing data, enhanced with very few dedicated \textit{HST} observations, significantly reducing both the time and resources invested. This is particularly meaningful in the era of large area surveys, with objects of interest typically observed by multiple surveys over several years. While the archival data alone is typically not sufficient to achieve good astrometric accuracy, supplementing it with few, well-planned observations can significantly improve the parallax and proper motion measurements. 

The 13 targets presented here populate the bottom of the main sequence. The color-magnitude diagrams of Figure~\ref{fig:CMD_full} and \ref{fig:CMD_HST_only} highlight the rapid change of the short-wavelength magnitudes as a function temperature, with Y dwarfs spanning a range of nearly 10 magnitudes in F110W absolute magnitude and 7.5 magnitudes in F110W--ch2 color when going from late T dwarfs ($\sim$500 K) to the coldest brown dwarf known (WISE~J0855; $\sim$250 K). Long-wavelength magnitudes evolve more slowly (changing by only $\sim$3 magnitudes over the same range of temperatures), but show a larger intrinsic scatter. Overall, the use of photometric distance and spectral type estimates becomes highly unreliable for Y dwarfs, emphasizing the need for astrometric and spectroscopic measurements.

With several dedicated efforts to characterize cold brown dwarfs in detail using JWST's exquisite photometric and spectroscopic capabilities, and with several ongoing surveys promising to discover more Y dwarfs, astrometry will continue to play a key role in the interpretation of these intriguing frigid objects.

\begin{acknowledgments}

We thank the anonymous referee for comments that significantly improved the quality of this manuscript. 
EC and RAMB acknowledge support from FONDECYT/ANID \# 124 0049. RAMB also acknowledges support from Fondo GEMINI, Astrónomo de Soporte GEMINI-ANID grant \# 3223 AS0002. 
This research is based in part on observations made with the NASA/ESA Hubble Space Telescope obtained from the Space Telescope Science Institute, which is operated by the Association of Universities for Research in Astronomy, Inc., under NASA contract NAS 5–26555. These observations are associated with program HST‐GO‐16243.
This work is based in part on observations made with the \textit{Spitzer} Space Telescope, which was operated by the Jet Propulsion Laboratory, California Institute of Technology under a contract with NASA. 
This publication makes use of data products from the Wide-field Infrared Survey Explorer, which was a joint project of the University of California, Los Angeles, and the Jet Propulsion Laboratory/California Institute of Technology, and NEOWISE, which was a joint project of the Jet Propulsion Laboratory/California Institute of Technology and the University of Arizona. WISE and NEOWISE were funded by the National Aeronautics and Space Administration.
\end{acknowledgments}

\vspace{5mm}
\facilities{HST (WFC3), Spitzer (IRAC), WISE, NEOWISE}

\software{IDL, python}

\bibliography{refs}{}

@ARTICLE{2010AJ....140.1868W,
  author =	 {{Wright}, E.~L. and {Eisenhardt}, P.~R.~M. and {Mainzer},
                  A.~K. and {Ressler}, M.~E. and {Cutri}, R.~M. and {Jarrett},
                  T. and {Kirkpatrick}, J.~D. and {Padgett}, D. and
                  {McMillan}, R.~S. and {Skrutskie}, M. and {Stanford},
                  S.~A. and {Cohen}, M. and {Walker}, R.~G. and {Mather},
                  J.~C. and {Leisawitz}, D. and {Gautier}, III, T.~N. and
                  {McLean}, I. and {Benford}, D. and {Lonsdale}, C.~J. and
                  {Blain}, A. and {Mendez}, B. and {Irace}, W.~R. and {Duval},
                  V. and {Liu}, F. and {Royer}, D. and {Heinrichsen}, I. and
                  {Howard}, J. and {Shannon}, M. and {Kendall}, M. and
                  {Walsh}, A.~L. and {Larsen}, M. and {Cardon}, J.~G. and
                  {Schick}, S. and {Schwalm}, M. and {Abid}, M. and
                  {Fabinsky}, B. and {Naes}, L. and {Tsai}, C.-W.},
  title =	 "{The Wide-field Infrared Survey Explorer (WISE): Mission
                  Description and Initial On-orbit Performance}",
  journal =	 {\aj},
  archivePrefix ="arXiv",
  eprint =	 {1008.0031},
  primaryClass = "astro-ph.IM",
  year =	 2010,
  month =	 dec,
  volume =	 140,
  pages =	 {1868-1881},
  doi =		 {10.1088/0004-6256/140/6/1868},
  adsurl =	 {http://cdsads.u-strasbg.fr/abs/2010AJ....140.1868W}
}

@ARTICLE{2011ApJ...743...50C,
  author =	 {{Cushing}, M.~C. and {Kirkpatrick}, J.~D. and {Gelino},
                  C.~R. and {Griffith}, R.~L. and {Skrutskie}, M.~F. and
                  {Mainzer}, A. and {Marsh}, K.~A. and {Beichman}, C.~A. and
                  {Burgasser}, A.~J. and {Prato}, L.~A. and {Simcoe},
                  R.~A. and {Marley}, M.~S. and {Saumon}, D. and {Freedman},
                  R.~S. and {Eisenhardt}, P.~R. and {Wright}, E.~L.  },
  title =	 "{The Discovery of Y Dwarfs using Data from the Wide-field
                  Infrared Survey Explorer (WISE)}",
  journal =	 {\apj},
  archivePrefix ="arXiv",
  eprint =	 {1108.4678},
  primaryClass = "astro-ph.SR",
  year =	 2011,
  month =	 dec,
  volume =	 743,
  eid =		 50,
  pages =	 50,
  doi =		 {10.1088/0004-637X/743/1/50},
  adsurl =	 {http://adsabs.harvard.edu/abs/2011ApJ...743...50C}
}

@ARTICLE{2012ApJ...753..156K,
  author =	 {{Kirkpatrick}, J.~D. and {Gelino}, C.~R. and {Cushing},
                  M.~C. and {Mace}, G.~N. and {Griffith}, R.~L. and
                  {Skrutskie}, M.~F. and {Marsh}, K.~A. and {Wright},
                  E.~L. and {Eisenhardt}, P.~R. and {McLean}, I.~S. and
                  {Mainzer}, A.~K. and {Burgasser}, A.~J. and {Tinney},
                  C.~G. and {Parker}, S. and {Salter}, G.},
  title =	 "{Further Defining Spectral Type ''Y'' and Exploring the
                  Low-mass End of the Field Brown Dwarf Mass Function}",
  journal =	 {\apj},
  archivePrefix ="arXiv",
  eprint =	 {1205.2122},
  primaryClass = "astro-ph.SR",
  year =	 2012,
  month =	 jul,
  volume =	 753,
  eid =		 156,
  pages =	 156,
  doi =		 {10.1088/0004-637X/753/2/156},
  adsurl =	 {http://cdsads.u-strasbg.fr/abs/2012ApJ...753..156K}
}

@ARTICLE{2013Sci...341.1492D,
  author =	 {{Dupuy}, T.~J. and {Kraus}, A.~L.},
  title =	 "{Distances, Luminosities, and Temperatures of the Coldest
                  Known Substellar Objects}",
  journal =	 {Science},
  archivePrefix ="arXiv",
  eprint =	 {1309.1422},
  primaryClass = "astro-ph.SR",
  year =	 2013,
  month =	 sep,
  volume =	 341,
  pages =	 {1492-1495},
  doi =		 {10.1126/science.1241917},
  adsurl =	 {http://adsabs.harvard.edu/abs/2013Sci...341.1492D}
}

@ARTICLE{2015ApJ...804...92S,
  author =	 {{Schneider}, A.~C. and {Cushing}, M.~C. and {Kirkpatrick},
                  J.~D. and {Gelino}, C.~R. and {Mace}, G.~N. and {Wright},
                  E.~L. and {Eisenhardt}, P.~R. and {Skrutskie}, M.~F. and
                  {Griffith}, R.~L. and {Marsh}, K.~A.},
  title =	 "{Hubble Space Telescope Spectroscopy of Brown Dwarfs
                  Discovered with the Wide-field Infrared Survey Explorer}",
  journal =	 {\apj},
  archivePrefix ="arXiv",
  eprint =	 {1502.05365},
  primaryClass = "astro-ph.SR",
  year =	 2015,
  month =	 may,
  volume =	 804,
  eid =		 92,
  pages =	 92,
  doi =		 {10.1088/0004-637X/804/2/92},
  adsurl =	 {http://cdsads.u-strasbg.fr/abs/2015ApJ...804...92S}
}

@ARTICLE{2014ApJ...786L..18L,
   author = {{Luhman}, K.~L.},
    title = "{Discovery of a \~{}250 K Brown Dwarf at 2 pc from the Sun}",
  journal = {\apjl},
archivePrefix = "arXiv",
   eprint = {1404.6501},
     year = 2014,
    month = may,
   volume = 786,
      eid = {L18},
    pages = {L18},
      doi = {10.1088/2041-8205/786/2/L18},
   adsurl = {http://adsabs.harvard.edu/abs/2014ApJ...786L..18L}
}

@ARTICLE{2019ApJS..240...19K,
   author = {{Kirkpatrick}, J.~D. and {Martin}, E.~C. and {Smart}, R.~L. and 
	{Cayago}, A.~J. and {Beichman}, C.~A. and {Marocco}, F. and 
	{Gelino}, C.~R. and {Faherty}, J.~K. and {Cushing}, M.~C. and 
	{Schneider}, A.~C. and {Mace}, G.~N. and {Tinney}, C.~G. and 
	{Wright}, E.~L. and {Lowrance}, P.~J. and {Ingalls}, J.~G. and 
	{Vrba}, F.~J. and {Munn}, J.~A. and {Dahm}, S.~E. and {McLean}, I.~S.
	},
    title = "{Preliminary Trigonometric Parallaxes of 184 Late-T and Y Dwarfs and an Analysis of the Field Substellar Mass Function into the {\ldquo}Planetary{\rdquo} Mass Regime}",
  journal = {\apjs},
archivePrefix = "arXiv",
   eprint = {1812.01208},
 primaryClass = "astro-ph.SR",
     year = 2019,
    month = feb,
   volume = 240,
      eid = {19},
    pages = {19},
      doi = {10.3847/1538-4365/aaf6af},
   adsurl = {http://adsabs.harvard.edu/abs/2019ApJS..240...19K}
}

@ARTICLE{2014AJ....148...82W,
   author = {{Wright}, E.~L. and {Mainzer}, A. and {Kirkpatrick}, J.~D. and 
	{Masci}, F. and {Cushing}, M.~C. and {Bauer}, J. and {Fajardo-Acosta}, S. and 
	{Gelino}, C.~R. and {Beichman}, C.~A. and {Skrutskie}, M.~F. and 
	{Grav}, T. and {Eisenhardt}, P.~R.~M. and {Cutri}, R.},
    title = "{NEOWISE-R Observation of the Coolest Known Brown Dwarf}",
  journal = {\aj},
archivePrefix = "arXiv",
   eprint = {1405.7350},
 primaryClass = "astro-ph.SR",
     year = 2014,
    month = nov,
   volume = 148,
      eid = {82},
    pages = {82},
      doi = {10.1088/0004-6256/148/5/82},
   adsurl = {http://adsabs.harvard.edu/abs/2014AJ....148...82W}
}

@ARTICLE{2019ApJ...881...17M,
       author = {{Marocco}, Federico and {Caselden}, Dan and {Meisner}, Aaron M. and
         {Kirkpatrick}, J. Davy and {Wright}, Edward L. and
         {Faherty}, Jacqueline K. and {Gelino}, Christopher R. and
         {Eisenhardt}, Peter R.~M. and {Fowler}, John W. and
         {Cushing}, Michael C. and {Cutri}, Roc M. and {Garcia}, Nelson and
         {Jarrett}, Thomas H. and {Koontz}, Renata and {Mainzer}, Amanda and
         {Marchese}, Elijah J. and {Mobasher}, Bahram and {Schlegel}, David J. and
         {Stern}, Daniel and {Teplitz}, Harry I.},
        title = "{CWISEP J193518.59─154620.3: An Extremely Cold Brown Dwarf in the Solar Neighborhood Discovered with CatWISE}",
      journal = {\apj},
         year = "2019",
        month = "Aug",
       volume = {881},
       number = {1},
          eid = {17},
        pages = {17},
          doi = {10.3847/1538-4357/ab2bf0},
archivePrefix = {arXiv},
       eprint = {1906.08913},
 primaryClass = {astro-ph.SR},
       adsurl = {https://ui.adsabs.harvard.edu/abs/2019ApJ...881...17M}
}

@ARTICLE{2020ApJ...889...74M,
       author = {{Meisner}, Aaron M. and {Caselden}, Dan and {Kirkpatrick}, J. Davy and
         {Marocco}, Federico and {Gelino}, Christopher R. and
         {Cushing}, Michael C. and {Eisenhardt}, Peter R.~M. and
         {Wright}, Edward L. and {Faherty}, Jacqueline K. and {Koontz}, Renata and
         {Marchese}, Elijah J. and {Khalil}, Mohammed and {Fowler}, John W. and
         {Schlafly}, Edward F.},
        title = "{Expanding the Y Dwarf Census with Spitzer Follow-up of the Coldest CatWISE Solar Neighborhood Discoveries}",
      journal = {\apj},
         year = 2020,
        month = feb,
       volume = {889},
       number = {2},
          eid = {74},
        pages = {74},
          doi = {10.3847/1538-4357/ab6215},
archivePrefix = {arXiv},
       eprint = {1911.12372},
 primaryClass = {astro-ph.SR},
       adsurl = {https://ui.adsabs.harvard.edu/abs/2020ApJ...889...74M}
}

@ARTICLE{2021ApJS..253....8M,
       author = {{Marocco}, Federico and {Eisenhardt}, Peter R.~M. and {Fowler}, John W. and {Kirkpatrick}, J. Davy and {Meisner}, Aaron M. and {Schlafly}, Edward F. and {Stanford}, S.~A. and {Garcia}, Nelson and {Caselden}, Dan and {Cushing}, Michael C. and {Cutri}, Roc M. and {Faherty}, Jacqueline K. and {Gelino}, Christopher R. and {Gonzalez}, Anthony H. and {Jarrett}, Thomas H. and {Koontz}, Renata and {Mainzer}, Amanda and {Marchese}, Elijah J. and {Mobasher}, Bahram and {Schlegel}, David J. and {Stern}, Daniel and {Teplitz}, Harry I. and {Wright}, Edward L.},
        title = "{The CatWISE2020 Catalog}",
      journal = {\apjs},
         year = 2021,
        month = mar,
       volume = {253},
       number = {1},
          eid = {8},
        pages = {8},
          doi = {10.3847/1538-4365/abd805},
archivePrefix = {arXiv},
       eprint = {2012.13084},
 primaryClass = {astro-ph.IM},
       adsurl = {https://ui.adsabs.harvard.edu/abs/2021ApJS..253....8M}
}

@ARTICLE{2020ApJ...899..123M,
       author = {{Meisner}, Aaron M. and {Faherty}, Jacqueline K. and {Kirkpatrick}, J. Davy and {Schneider}, Adam C. and {Caselden}, Dan and {Gagn{\'e}}, Jonathan and {Kuchner}, Marc J. and {Burgasser}, Adam J. and {Casewell}, Sarah L. and {Debes}, John H. and {Artigau}, {\'E}tienne and {Bardalez Gagliuffi}, Daniella C. and {Logsdon}, Sarah E. and {Kiman}, Rocio and {Allers}, Katelyn and {Hsu}, Chih-chun and {Wisniewski}, John P. and {Allen}, Michaela B. and {Beaulieu}, Paul and {Colin}, Guillaume and {Durantini Luca}, Hugo A. and {Goodman}, Sam and {Gramaize}, L{\'e}opold and {Hamlet}, Leslie K. and {Hinckley}, Ken and {Kiwy}, Frank and {Martin}, David W. and {Pendrill}, William and {Rothermich}, Austin and {Sainio}, Arttu and {Sch{\"u}mann}, J{\"o}rg and {Andersen}, Nikolaj Stevnbak and {Tanner}, Christopher and {Thakur}, Vinod and {Th{\'e}venot}, Melina and {Walla}, Jim and {W{\k{e}}dracki}, Zbigniew and {Aganze}, Christian and {Gerasimov}, Roman and {Theissen}, Christopher and {Backyard Worlds: Planet 9 Collaboration}},
        title = "{Spitzer Follow-up of Extremely Cold Brown Dwarfs Discovered by the Backyard Worlds: Planet 9 Citizen Science Project}",
      journal = {\apj},
         year = 2020,
        month = aug,
       volume = {899},
       number = {2},
          eid = {123},
        pages = {123},
          doi = {10.3847/1538-4357/aba633},
archivePrefix = {arXiv},
       eprint = {2008.06396},
 primaryClass = {astro-ph.SR},
       adsurl = {https://ui.adsabs.harvard.edu/abs/2020ApJ...899..123M}
}

@ARTICLE{2021ApJS..253....7K,
       author = {{Kirkpatrick}, J. Davy and {Gelino}, Christopher R. and {Faherty}, Jacqueline K. and {Meisner}, Aaron M. and {Caselden}, Dan and {Schneider}, Adam C. and {Marocco}, Federico and {Cayago}, Alfred J. and {Smart}, R.~L. and {Eisenhardt}, Peter R. and {Kuchner}, Marc J. and {Wright}, Edward L. and {Cushing}, Michael C. and {Allers}, Katelyn N. and {Bardalez Gagliuffi}, Daniella C. and {Burgasser}, Adam J. and {Gagn{\'e}}, Jonathan and {Logsdon}, Sarah E. and {Martin}, Emily C. and {Ingalls}, James G. and {Lowrance}, Patrick J. and {Abrahams}, Ellianna S. and {Aganze}, Christian and {Gerasimov}, Roman and {Gonzales}, Eileen C. and {Hsu}, Chih-Chun and {Kamraj}, Nikita and {Kiman}, Rocio and {Rees}, Jon and {Theissen}, Christopher and {Ammar}, Kareem and {Andersen}, Nikolaj Stevnbak and {Beaulieu}, Paul and {Colin}, Guillaume and {Elachi}, Charles A. and {Goodman}, Samuel J. and {Gramaize}, L{\'e}opold and {Hamlet}, Leslie K. and {Hong}, Justin and {Jonkeren}, Alexander and {Khalil}, Mohammed and {Martin}, David W. and {Pendrill}, William and {Pumphrey}, Benjamin and {Rothermich}, Austin and {Sainio}, Arttu and {Stenner}, Andres and {Tanner}, Christopher and {Th{\'e}venot}, Melina and {Voloshin}, Nikita V. and {Walla}, Jim and {W{\k{e}}dracki}, Zbigniew and {Backyard Worlds: Planet 9 Collaboration}},
        title = "{The Field Substellar Mass Function Based on the Full-sky 20 pc Census of 525 L, T, and Y Dwarfs}",
      journal = {\apjs},
         year = 2021,
        month = mar,
       volume = {253},
       number = {1},
          eid = {7},
        pages = {7},
          doi = {10.3847/1538-4365/abd107},
archivePrefix = {arXiv},
       eprint = {2011.11616},
 primaryClass = {astro-ph.SR},
       adsurl = {https://ui.adsabs.harvard.edu/abs/2021ApJS..253....7K}
}

@ARTICLE{2023AJ....165...36M,
       author = {{Meisner}, Aaron M. and {Caselden}, Dan and {Schlafly}, Edward F. and {Kiwy}, Frank},
        title = "{unTimely: a Full-sky, Time-domain unWISE Catalog}",
      journal = {\aj},
         year = 2023,
        month = feb,
       volume = {165},
       number = {2},
          eid = {36},
        pages = {36},
          doi = {10.3847/1538-3881/aca2ab},
archivePrefix = {arXiv},
       eprint = {2209.14327},
 primaryClass = {astro-ph.IM},
       adsurl = {https://ui.adsabs.harvard.edu/abs/2023AJ....165...36M}
}

@ARTICLE{2014ApJ...792...30M,
       author = {{Mainzer}, A. and {Bauer}, J. and {Cutri}, R.~M. and {Grav}, T. and {Masiero}, J. and {Beck}, R. and {Clarkson}, P. and {Conrow}, T. and {Dailey}, J. and {Eisenhardt}, P. and {Fabinsky}, B. and {Fajardo-Acosta}, S. and {Fowler}, J. and {Gelino}, C. and {Grillmair}, C. and {Heinrichsen}, I. and {Kendall}, M. and {Kirkpatrick}, J. Davy and {Liu}, F. and {Masci}, F. and {McCallon}, H. and {Nugent}, C.~R. and {Papin}, M. and {Rice}, E. and {Royer}, D. and {Ryan}, T. and {Sevilla}, P. and {Sonnett}, S. and {Stevenson}, R. and {Thompson}, D.~B. and {Wheelock}, S. and {Wiemer}, D. and {Wittman}, M. and {Wright}, E. and {Yan}, L.},
        title = "{Initial Performance of the NEOWISE Reactivation Mission}",
      journal = {\apj},
         year = 2014,
        month = sep,
       volume = {792},
       number = {1},
          eid = {30},
        pages = {30},
          doi = {10.1088/0004-637X/792/1/30},
archivePrefix = {arXiv},
       eprint = {1406.6025},
 primaryClass = {astro-ph.EP},
       adsurl = {https://ui.adsabs.harvard.edu/abs/2014ApJ...792...30M}
}

@ARTICLE{2024ApJS..271...55K,
       author = {{Kirkpatrick}, J. Davy and {Marocco}, Federico and {Gelino}, Christopher R. and {Raghu}, Yadukrishna and {Faherty}, Jacqueline K. and {Bardalez Gagliuffi}, Daniella C. and {Schurr}, Steven D. and {Apps}, Kevin and {Schneider}, Adam C. and {Meisner}, Aaron M. and {Kuchner}, Marc J. and {Caselden}, Dan and {Smart}, R.~L. and {Casewell}, S.~L. and {Raddi}, Roberto and {Kesseli}, Aurora and {Stevnbak Andersen}, Nikolaj and {Antonini}, Edoardo and {Beaulieu}, Paul and {Bickle}, Thomas P. and {Bilsing}, Martin and {Chieng}, Raymond and {Colin}, Guillaume and {Deen}, Sam and {Dereveanco}, Alexandru and {Doll}, Katharina and {Durantini Luca}, Hugo A. and {Frazer}, Anya and {Gantier}, Jean Marc and {Gramaize}, L{\'e}opold and {Grant}, Kristin and {Hamlet}, Leslie K. and {Higashimura}, Hiro and {Hyogo}, Michiharu and {Ja{\l}owiczor}, Peter A. and {Jonkeren}, Alexander and {Kabatnik}, Martin and {Kiwy}, Frank and {Martin}, David W. and {Michaels}, Marianne N. and {Pendrill}, William and {Pessanha Machado}, Celso and {Pumphrey}, Benjamin and {Rothermich}, Austin and {Russwurm}, Rebekah and {Sainio}, Arttu and {Sanchez}, John and {Sapelkin-Tambling}, Fyodor Theo and {Sch{\"u}mann}, J{\"o}rg and {Selg-Mann}, Karl and {Singh}, Harshdeep and {Stenner}, Andres and {Sun}, Guoyou and {Tanner}, Christopher and {Th{\'e}venot}, Melina and {Ventura}, Maurizio and {Voloshin}, Nikita V. and {Walla}, Jim and {W{\k{e}}dracki}, Zbigniew and {Adorno}, Jose I. and {Aganze}, Christian and {Allers}, Katelyn N. and {Brooks}, Hunter and {Burgasser}, Adam J. and {Calamari}, Emily and {Connor}, Thomas and {Costa}, Edgardo and {Eisenhardt}, Peter R. and {Gagn{\'e}}, Jonathan and {Gerasimov}, Roman and {Gonzales}, Eileen C. and {Hsu}, Chih-Chun and {Kiman}, Rocio and {Li}, Guodong and {Low}, Ryan and {Mamajek}, Eric and {Pantoja}, Blake M. and {Popinchalk}, Mark and {Rees}, Jon M. and {Stern}, Daniel and {Su{\'a}rez}, Genaro and {Theissen}, Christopher and {Tsai}, Chao-Wei and {Vos}, Johanna M. and {Zurek}, David and {The Backyard Worlds: Planet 9 Collaboration}},
        title = "{The Initial Mass Function Based on the Full-sky 20 pc Census of {\ensuremath{\sim}}3600 Stars and Brown Dwarfs}",
      journal = {\apjs},
         year = 2024,
        month = apr,
       volume = {271},
       number = {2},
          eid = {55},
        pages = {55},
          doi = {10.3847/1538-4365/ad24e2},
archivePrefix = {arXiv},
       eprint = {2312.03639},
 primaryClass = {astro-ph.SR},
       adsurl = {https://ui.adsabs.harvard.edu/abs/2024ApJS..271...55K}
}

@ARTICLE{2020ApJ...888L..19M,
       author = {{Marocco}, Federico and {Kirkpatrick}, J. Davy and {Meisner}, Aaron M. and {Caselden}, Dan and {Eisenhardt}, Peter R.~M. and {Cushing}, Michael C. and {Faherty}, Jacqueline K. and {Gelino}, Christopher R. and {Wright}, Edward L.},
        title = "{Improved Infrared Photometry and a Preliminary Parallax Measurement for the Extremely Cold Brown Dwarf CWISEP J144606.62-231717.8}",
      journal = {\apjl},
         year = 2020,
        month = jan,
       volume = {888},
       number = {2},
          eid = {L19},
        pages = {L19},
          doi = {10.3847/2041-8213/ab6201},
archivePrefix = {arXiv},
       eprint = {1912.07692},
 primaryClass = {astro-ph.SR},
       adsurl = {https://ui.adsabs.harvard.edu/abs/2020ApJ...888L..19M}
}

@ARTICLE{1973PASP...85..573L,
       author = {{Lutz}, Thomas E. and {Kelker}, Douglas H.},
        title = "{On the Use of Trigonometric Parallaxes for the Calibration of Luminosity Systems: Theory}",
      journal = {\pasp},
         year = 1973,
        month = oct,
       volume = {85},
       number = {507},
        pages = {573},
          doi = {10.1086/129506},
       adsurl = {https://ui.adsabs.harvard.edu/abs/1973PASP...85..573L}
}

@ARTICLE{2020MNRAS.494.2068B,
       author = {{Bedin}, L.~R. and {Fontanive}, C.},
        title = "{Extending Gaia DR2 with HST narrow-field astrometry - II. Refining the method on WISE J163940.83-684738.6}",
      journal = {\mnras},
         year = 2020,
        month = may,
       volume = {494},
       number = {2},
        pages = {2068-2075},
          doi = {10.1093/mnras/staa540},
archivePrefix = {arXiv},
       eprint = {2002.09331},
 primaryClass = {astro-ph.SR},
       adsurl = {https://ui.adsabs.harvard.edu/abs/2020MNRAS.494.2068B}
}

@ARTICLE{2017AJ....153...14W,
       author = {{Winters}, Jennifer G. and {Sevrinsky}, R. Andrew and {Jao}, Wei-Chun and {Henry}, Todd J. and {Riedel}, Adric R. and {Subasavage}, John P. and {Lurie}, John C. and {Ianna}, Philip A. and {Finch}, Charlie T.},
        title = "{The Solar Neighborhood XXXVIII. Results from the CTIO/SMARTS 0.9m: Trigonometric Parallaxes for 151 Nearby M Dwarf Systems}",
      journal = {\aj},
         year = 2017,
        month = jan,
       volume = {153},
       number = {1},
          eid = {14},
        pages = {14},
          doi = {10.3847/1538-3881/153/1/14},
archivePrefix = {arXiv},
       eprint = {1610.07552},
 primaryClass = {astro-ph.SR},
       adsurl = {https://ui.adsabs.harvard.edu/abs/2017AJ....153...14W}
}

@ARTICLE{2021AJ....161...42B,
       author = {{Best}, William M.~J. and {Liu}, Michael C. and {Magnier}, Eugene A. and {Dupuy}, Trent J.},
        title = "{A Volume-limited Sample of Ultracool Dwarfs. I. Construction, Space Density, and a Gap in the L/T Transition}",
      journal = {\aj},
         year = 2021,
        month = jan,
       volume = {161},
       number = {1},
          eid = {42},
        pages = {42},
          doi = {10.3847/1538-3881/abc893},
archivePrefix = {arXiv},
       eprint = {2010.15853},
 primaryClass = {astro-ph.SR},
       adsurl = {https://ui.adsabs.harvard.edu/abs/2021AJ....161...42B}
}

@ARTICLE{2018MNRAS.481.3548S,
       author = {{Smart}, R.~L. and {Bucciarelli}, B. and {Jones}, H.~R.~A. and {Marocco}, F. and {Andrei}, A.~H. and {Goldman}, B. and {Mendez}, R.~A. and {d'Avila}, V.~A. and {Burningham}, B. and {Camargo}, J.~I.~B. and {Crosta}, M.~T. and {Dapr{\`a}}, M. and {Jenkins}, J.~S. and {Lachaume}, R. and {Lattanzi}, M.~G. and {Penna}, J.~L. and {Pinfield}, D.~J. and {da Silva Neto}, D.~N. and {Sozzetti}, A. and {Vecchiato}, A.},
        title = "{Parallaxes of Southern Extremely Cool objects III: 118 L and T dwarfs}",
      journal = {\mnras},
         year = 2018,
        month = dec,
       volume = {481},
       number = {3},
        pages = {3548-3562},
          doi = {10.1093/mnras/sty2520},
archivePrefix = {arXiv},
       eprint = {1811.00672},
 primaryClass = {astro-ph.SR},
       adsurl = {https://ui.adsabs.harvard.edu/abs/2018MNRAS.481.3548S}
}

@ARTICLE{2023A&A...674A...1G,
       author = {{Gaia Collaboration} and {Vallenari}, A. and {Brown}, A.~G.~A. and {Prusti}, T. and {de Bruijne}, J.~H.~J. and {Arenou}, F. and {Babusiaux}, C. and {Biermann}, M. and {Creevey}, O.~L. and {Ducourant}, C. and {Evans}, D.~W. and {Eyer}, L. and {Guerra}, R. and {Hutton}, A. and {Jordi}, C. and {Klioner}, S.~A. and {Lammers}, U.~L. and {Lindegren}, L. and {Luri}, X. and {Mignard}, F. and {Panem}, C. and {Pourbaix}, D. and {Randich}, S. and {Sartoretti}, P. and {Soubiran}, C. and {Tanga}, P. and {Walton}, N.~A. and {Bailer-Jones}, C.~A.~L. and {Bastian}, U. and {Drimmel}, R. and {Jansen}, F. and {Katz}, D. and {Lattanzi}, M.~G. and {van Leeuwen}, F. and {Bakker}, J. and {Cacciari}, C. and {Casta{\~n}eda}, J. and {De Angeli}, F. and {Fabricius}, C. and {Fouesneau}, M. and {Fr{\'e}mat}, Y. and {Galluccio}, L. and {Guerrier}, A. and {Heiter}, U. and {Masana}, E. and {Messineo}, R. and {Mowlavi}, N. and {Nicolas}, C. and {Nienartowicz}, K. and {Pailler}, F. and {Panuzzo}, P. and {Riclet}, F. and {Roux}, W. and {Seabroke}, G.~M. and {Sordo}, R. and {Th{\'e}venin}, F. and {Gracia-Abril}, G. and {Portell}, J. and {Teyssier}, D. and {Altmann}, M. and {Andrae}, R. and {Audard}, M. and {Bellas-Velidis}, I. and {Benson}, K. and {Berthier}, J. and {Blomme}, R. and {Burgess}, P.~W. and {Busonero}, D. and {Busso}, G. and {C{\'a}novas}, H. and {Carry}, B. and {Cellino}, A. and {Cheek}, N. and {Clementini}, G. and {Damerdji}, Y. and {Davidson}, M. and {de Teodoro}, P. and {Nu{\~n}ez Campos}, M. and {Delchambre}, L. and {Dell'Oro}, A. and {Esquej}, P. and {Fern{\'a}ndez-Hern{\'a}ndez}, J. and {Fraile}, E. and {Garabato}, D. and {Garc{\'\i}a-Lario}, P. and {Gosset}, E. and {Haigron}, R. and {Halbwachs}, J. -L. and {Hambly}, N.~C. and {Harrison}, D.~L. and {Hern{\'a}ndez}, J. and {Hestroffer}, D. and {Hodgkin}, S.~T. and {Holl}, B. and {Jan{\ss}en}, K. and {Jevardat de Fombelle}, G. and {Jordan}, S. and {Krone-Martins}, A. and {Lanzafame}, A.~C. and {L{\"o}ffler}, W. and {Marchal}, O. and {Marrese}, P.~M. and {Moitinho}, A. and {Muinonen}, K. and {Osborne}, P. and {Pancino}, E. and {Pauwels}, T. and {Recio-Blanco}, A. and {Reyl{\'e}}, C. and {Riello}, M. and {Rimoldini}, L. and {Roegiers}, T. and {Rybizki}, J. and {Sarro}, L.~M. and {Siopis}, C. and {Smith}, M. and {Sozzetti}, A. and {Utrilla}, E. and {van Leeuwen}, M. and {Abbas}, U. and {{\'A}brah{\'a}m}, P. and {Abreu Aramburu}, A. and {Aerts}, C. and {Aguado}, J.~J. and {Ajaj}, M. and {Aldea-Montero}, F. and {Altavilla}, G. and {{\'A}lvarez}, M.~A. and {Alves}, J. and {Anders}, F. and {Anderson}, R.~I. and {Anglada Varela}, E. and {Antoja}, T. and {Baines}, D. and {Baker}, S.~G. and {Balaguer-N{\'u}{\~n}ez}, L. and {Balbinot}, E. and {Balog}, Z. and {Barache}, C. and {Barbato}, D. and {Barros}, M. and {Barstow}, M.~A. and {Bartolom{\'e}}, S. and {Bassilana}, J. -L. and {Bauchet}, N. and {Becciani}, U. and {Bellazzini}, M. and {Berihuete}, A. and {Bernet}, M. and {Bertone}, S. and {Bianchi}, L. and {Binnenfeld}, A. and {Blanco-Cuaresma}, S. and {Blazere}, A. and {Boch}, T. and {Bombrun}, A. and {Bossini}, D. and {Bouquillon}, S. and {Bragaglia}, A. and {Bramante}, L. and {Breedt}, E. and {Bressan}, A. and {Brouillet}, N. and {Brugaletta}, E. and {Bucciarelli}, B. and {Burlacu}, A. and {Butkevich}, A.~G. and {Buzzi}, R. and {Caffau}, E. and {Cancelliere}, R. and {Cantat-Gaudin}, T. and {Carballo}, R. and {Carlucci}, T. and {Carnerero}, M.~I. and {Carrasco}, J.~M. and {Casamiquela}, L. and {Castellani}, M. and {Castro-Ginard}, A. and {Chaoul}, L. and {Charlot}, P. and {Chemin}, L. and {Chiaramida}, V. and {Chiavassa}, A. and {Chornay}, N. and {Comoretto}, G. and {Contursi}, G. and {Cooper}, W.~J. and {Cornez}, T. and {Cowell}, S. and {Crifo}, F. and {Cropper}, M. and {Crosta}, M. and {Crowley}, C. and {Dafonte}, C. and {Dapergolas}, A. and {David}, M. and {David}, P. and {de Laverny}, P. and {De Luise}, F. and {De March}, R.},
        title = "{Gaia Data Release 3. Summary of the content and survey properties}",
      journal = {\aap},
         year = 2023,
        month = jun,
       volume = {674},
          eid = {A1},
        pages = {A1},
          doi = {10.1051/0004-6361/202243940},
archivePrefix = {arXiv},
       eprint = {2208.00211},
 primaryClass = {astro-ph.GA},
       adsurl = {https://ui.adsabs.harvard.edu/abs/2023A&A...674A...1G}
}

@ARTICLE{1955ApJ...121..161S,
       author = {{Salpeter}, Edwin E.},
        title = "{The Luminosity Function and Stellar Evolution.}",
      journal = {\apj},
         year = 1955,
        month = jan,
       volume = {121},
        pages = {161},
          doi = {10.1086/145971},
       adsurl = {https://ui.adsabs.harvard.edu/abs/1955ApJ...121..161S}
}

@ARTICLE{1968ApJ...151..393S,
       author = {{Schmidt}, Maarten},
        title = "{Space Distribution and Luminosity Functions of Quasi-Stellar Radio Sources}",
      journal = {\apj},
         year = 1968,
        month = feb,
       volume = {151},
        pages = {393},
          doi = {10.1086/149446},
       adsurl = {https://ui.adsabs.harvard.edu/abs/1968ApJ...151..393S}
}

@ARTICLE{2024ApJ...973..107B,
       author = {{Beiler}, Samuel A. and {Cushing}, Michael C. and {Kirkpatrick}, J. Davy and {Schneider}, Adam C. and {Mukherjee}, Sagnick and {Marley}, Mark S. and {Marocco}, Federico and {Smart}, Richard L.},
        title = "{Precise Bolometric Luminosities and Effective Temperatures of 23 Late-T and Y Dwarfs Obtained with JWST}",
      journal = {\apj},
         year = 2024,
        month = oct,
       volume = {973},
       number = {2},
          eid = {107},
        pages = {107},
          doi = {10.3847/1538-4357/ad6301},
archivePrefix = {arXiv},
       eprint = {2407.08518},
 primaryClass = {astro-ph.SR},
       adsurl = {https://ui.adsabs.harvard.edu/abs/2024ApJ...973..107B}
}

@ARTICLE{2018ApJ...862..173T,
       author = {{Theissen}, Christopher A.},
        title = "{Parallaxes of Cool Objects with WISE: Filling in for Gaia}",
      journal = {\apj},
         year = 2018,
        month = aug,
       volume = {862},
       number = {2},
          eid = {173},
        pages = {173},
          doi = {10.3847/1538-4357/aaccfa},
archivePrefix = {arXiv},
       eprint = {1710.11127},
 primaryClass = {astro-ph.SR},
       adsurl = {https://ui.adsabs.harvard.edu/abs/2018ApJ...862..173T}
}

@ARTICLE{2021ApJ...920...20C,
       author = {{Cushing}, Michael C. and {Schneider}, Adam C. and {Kirkpatrick}, J. Davy and {Morley}, Caroline V. and {Marley}, Mark S. and {Gelino}, Christopher R. and {Mace}, Gregory N. and {Wright}, Edward L. and {Eisenhardt}, Peter R. and {Skrutskie}, Michael F. and {Marsh}, Kenneth A.},
        title = "{An Improved Near-infrared Spectrum of the Archetype Y Dwarf WISEP J182831.08+265037.8}",
      journal = {\apj},
         year = 2021,
        month = oct,
       volume = {920},
       number = {1},
          eid = {20},
        pages = {20},
          doi = {10.3847/1538-4357/ac12cb},
archivePrefix = {arXiv},
       eprint = {2107.00506},
 primaryClass = {astro-ph.SR},
       adsurl = {https://ui.adsabs.harvard.edu/abs/2021ApJ...920...20C}
}

@ARTICLE{2014AJ....147..113C,
       author = {{Cushing}, Michael C. and {Kirkpatrick}, J. Davy and {Gelino}, Christopher R. and {Mace}, Gregory N. and {Skrutskie}, Michael F. and {Gould}, Andrew},
        title = "{Three New Cool Brown Dwarfs Discovered with the Wide-field Infrared Survey Explorer (WISE) and an Improved Spectrum of the Y0 Dwarf WISE J041022.71+150248.4}",
      journal = {\aj},
         year = 2014,
        month = may,
       volume = {147},
       number = {5},
          eid = {113},
        pages = {113},
          doi = {10.1088/0004-6256/147/5/113},
archivePrefix = {arXiv},
       eprint = {1402.1378},
 primaryClass = {astro-ph.SR},
       adsurl = {https://ui.adsabs.harvard.edu/abs/2014AJ....147..113C}
}

@ARTICLE{2016ApJ...823L..35S,
       author = {{Schneider}, Adam C. and {Cushing}, Michael C. and {Kirkpatrick}, J. Davy and {Gelino}, Christopher R.},
        title = "{The Collapse of the Wien Tail in the Coldest Brown Dwarf? Hubble Space Telescope Near-infrared Photometry of WISE J085510.83-071442.5}",
      journal = {\apjl},
         year = 2016,
        month = jun,
       volume = {823},
       number = {2},
          eid = {L35},
        pages = {L35},
          doi = {10.3847/2041-8205/823/2/L35},
archivePrefix = {arXiv},
       eprint = {1605.05618},
 primaryClass = {astro-ph.EP},
       adsurl = {https://ui.adsabs.harvard.edu/abs/2016ApJ...823L..35S}
}

@ARTICLE{2016AJ....152...78L,
       author = {{Luhman}, K.~L. and {Esplin}, T.~L.},
        title = "{The Spectral Energy Distribution of the Coldest Known Brown Dwarf}",
      journal = {\aj},
         year = 2016,
        month = sep,
       volume = {152},
       number = {3},
          eid = {78},
        pages = {78},
          doi = {10.3847/0004-6256/152/3/78},
archivePrefix = {arXiv},
       eprint = {1605.06655},
 primaryClass = {astro-ph.SR},
       adsurl = {https://ui.adsabs.harvard.edu/abs/2016AJ....152...78L}
}

@INPROCEEDINGS{2008SPIE.7010E..1EK,
       author = {{Kimble}, Randy A. and {MacKenty}, John W. and {O'Connell}, Robert W. and {Townsend}, Jacqueline A.},
        title = "{Wide Field Camera 3: a powerful new imager for the Hubble Space Telescope}",
    booktitle = {Space Telescopes and Instrumentation 2008: Optical, Infrared, and Millimeter},
         year = 2008,
       editor = {{Oschmann}, Jr., Jacobus M. and {de Graauw}, Mattheus W.~M. and {MacEwen}, Howard A.},
       series = {Society of Photo-Optical Instrumentation Engineers (SPIE) Conference Series},
       volume = {7010},
        month = jul,
          eid = {70101E},
        pages = {70101E},
          doi = {10.1117/12.789581},
       adsurl = {https://ui.adsabs.harvard.edu/abs/2008SPIE.7010E..1EK}
}

@BOOK{1997ilt..book.....R,
       author = {{Rodr{\'\i}guez Espinosa}, Jos{\'e} M. and {Herrero}, A. and {S{\'a}nchez}, F.},
        title = "{Instrumentation for large telescopes : VII Canary Islands Winter School of Astrophysics}",
         year = 1997,
       adsurl = {https://ui.adsabs.harvard.edu/abs/1997ilt..book.....R}
}

@ARTICLE{1985MNRAS.214..575I,
       author = {{Irwin}, M.~J.},
        title = "{Automatic analysis of crowded fields.}",
      journal = {\mnras},
         year = 1985,
        month = jun,
       volume = {214},
        pages = {575-604},
          doi = {10.1093/mnras/214.4.575},
       adsurl = {https://ui.adsabs.harvard.edu/abs/1985MNRAS.214..575I}
}

@INPROCEEDINGS{2009ASPC..411..251M,
       author = {{Markwardt}, C.~B.},
        title = "{Non-linear Least-squares Fitting in IDL with MPFIT}",
    booktitle = {Astronomical Data Analysis Software and Systems XVIII},
         year = 2009,
       editor = {{Bohlender}, D.~A. and {Durand}, D. and {Dowler}, P.},
       series = {Astronomical Society of the Pacific Conference Series},
       volume = {411},
        month = sep,
        pages = {251},
          doi = {10.48550/arXiv.0902.2850},
archivePrefix = {arXiv},
       eprint = {0902.2850},
 primaryClass = {astro-ph.IM},
       adsurl = {https://ui.adsabs.harvard.edu/abs/2009ASPC..411..251M}
}

@TECHREPORT{2021jwst.rept.7716A,
       author = {{Anderson}, Jay and {Fall}, S. Michael and {Astrometry Working Group}},
        title = "{The JWST Calibration Field: Absolute Astrometry and Proper Motions with GAIA and a Second HST Epoch}",
  institution = {STScI},
         year = 2021,
       number = {Technical Report JWST-STScI-007716},
 howpublished = {Technical Report JWST-STScI-007716},
       adsurl = {https://ui.adsabs.harvard.edu/abs/2021jwst.rept.7716A}
}
\bibliographystyle{aasjournalv7}


\end{document}